\documentclass[preprint,12pt,authoryear]{elsarticle}
\usepackage[utf8]{inputenc}
\usepackage[T1]{fontenc}
\usepackage{lmodern}
\usepackage{textcomp}
\usepackage{amsmath,amssymb}
\usepackage{graphicx}
\usepackage{booktabs,tabularx,array,xltabular}
\usepackage[table]{xcolor}
\usepackage{newunicodechar}
\usepackage[hidelinks]{hyperref}
\graphicspath{{./}}
\newunicodechar{−}{\ensuremath{-}}
\newunicodechar{×}{\ensuremath{\times}}
\newunicodechar{±}{\ensuremath{\pm}}
\newunicodechar{÷}{\ensuremath{\div}}
\newunicodechar{≥}{\ensuremath{\geq}}
\newunicodechar{≤}{\ensuremath{\leq}}
\newunicodechar{·}{\ensuremath{\cdot}}
\newunicodechar{²}{\textsuperscript{2}}
\newunicodechar{³}{\textsuperscript{3}}
\newunicodechar{⁵}{\textsuperscript{5}}
\newunicodechar{¹}{\textsuperscript{1}}
\newunicodechar{⁻}{\textsuperscript{\ensuremath{-}}}
\newunicodechar{°}{\textdegree}
\newunicodechar{√}{\ensuremath{\surd}}
\newunicodechar{≈}{\ensuremath{\approx}}
\newunicodechar{→}{\ensuremath{\rightarrow}}
\makeatletter
\def\ps@pprintTitle{\let\@oddhead\@empty\let\@evenhead\@empty\def\@oddfoot{\footnotesize\itshape Preprint\hfill\today}\let\@evenfoot\@oddfoot}
\makeatother

\begin{document}
\begin{frontmatter}
\title{LAaDT-DSS: decision support for non-experts on rooftop PV, incentive and grid simulations under Turkish tariff rules}
\author[su]{Baran Kaynak\corref{cor1}}
\ead{kaynak@sakarya.edu.tr}
\ead[url]{https://orcid.org/0000-0002-9004-2639}
\cortext[cor1]{Corresponding author.}
\address[su]{Department of Information Systems Engineering, Sakarya University, Sakarya, T\"urkiye}
\begin{abstract}
Rooftop photovoltaics raise three questions: should a household invest, which incentive costs the state least, and does the feeder stay within limits? Those who ask are rarely modellers. We present LAaDT-DSS, decision-support software that answers all three under Turkish tariff rules from one simulation core. Non-experts reach it by clicking a house on a 3D map or by asking in plain words; a language model is asked to call the simulation tools, not to compute. Each entry of its 125-entry parameter registry is marked as sourced or estimated. Case studies test the system. Its PV yield lies within ±3.6~\% of two independent tools at equal losses. Across 2,048 draws of its estimated inputs, the answer for PV alone held in all but one; with a battery, it changed in 31–41~\% of the draws. Four scripted chat scenarios ran end to end; no user study was run.
\end{abstract}

\end{frontmatter}

\noindent\fbox{\parbox{\dimexpr\textwidth-2\fboxsep-2\fboxrule\relax}{\small\itshape This manuscript is an arXiv preprint and has been submitted for publication in a peer-reviewed journal. It has not yet been peer reviewed. Later versions may differ in content.}}
\vspace{0.5em}

\section{Introduction}
Rooftop photovoltaics (PV) put urban roofs to low-carbon use, and three parties decide on them. A household asks whether a system pays for itself. The state asks which incentive makes it pay at the least cost to the budget. The distribution utility asks whether its feeders stay within their voltage and loading limits. Since a 2019 decree, Turkish households net their imports and exports per month (\citealp{cecen2022}), a rule the 2026 amendment kept for the first ten years (\citealp{epdk2026a}). Under the earlier feed-in tariff, the household's answer depended on the incentive and the region (\citealp{duman2020}), and the state has several incentive designs to choose from (\citealp{cecen2022}).

A decision tool gives each party one answer. Behind it sit many inputs that nobody has measured for the case at hand, such as costs, discount rates and network details, which the person asking cannot test. Sensitivity analysis addresses this: \citet{saltelli2019} set out how to do it well, and \citet{razavi2021} how it supports decisions. Related measures bound the share of a parameter space in which a decision cannot change (\citealp{ravalico2010}) or count the scenarios in which a strategy performs adequately (\citealp{maier2016}). We report how often an answer changes: the share of draws, within the declared input ranges, in which it differs from the answer with the registry values.

In the studies we reviewed (Section~2), no single tool gives the household, state and utility answers together, is built for people who are not modellers, compares its parts with independent tools and reports how often each answer changes. Without these checks, a user cannot tell which answers to trust.

We present LAaDT-DSS, a decision-support system that gives all three answers from one simulation core and one parameter registry (Fig.~1). Every entry of the registry is marked as sourced or estimated, and 99 of its 125 entries are estimates. The platform works at two levels: one building at a time, and a whole district screened with shortcuts. The people who ask are rarely modellers, so it is built for use without knowing the core. They reach it through a map studio, where a click on a house sizes a system, or through a chat in plain words. Access is a design aim. Four scripted scenarios demonstrate the chat; neither route was tested with users (Section~3.4). The chat follows a published pattern, in which a language model is asked to call the simulation tools, not to compute (\citealp{mylonas2026}; \citealp{zhang2025}). We test the system with four questions. How far do its parts lie from independent tools (RQ1)? How far do the district shortcuts lie from the per-building model (RQ2)? How often does each answer change under the declared ranges of its estimates (RQ3)? And what do the three answers look like for one real district (RQ4)?

\begin{figure}[tbp]
\centering
\includegraphics[width=\textwidth]{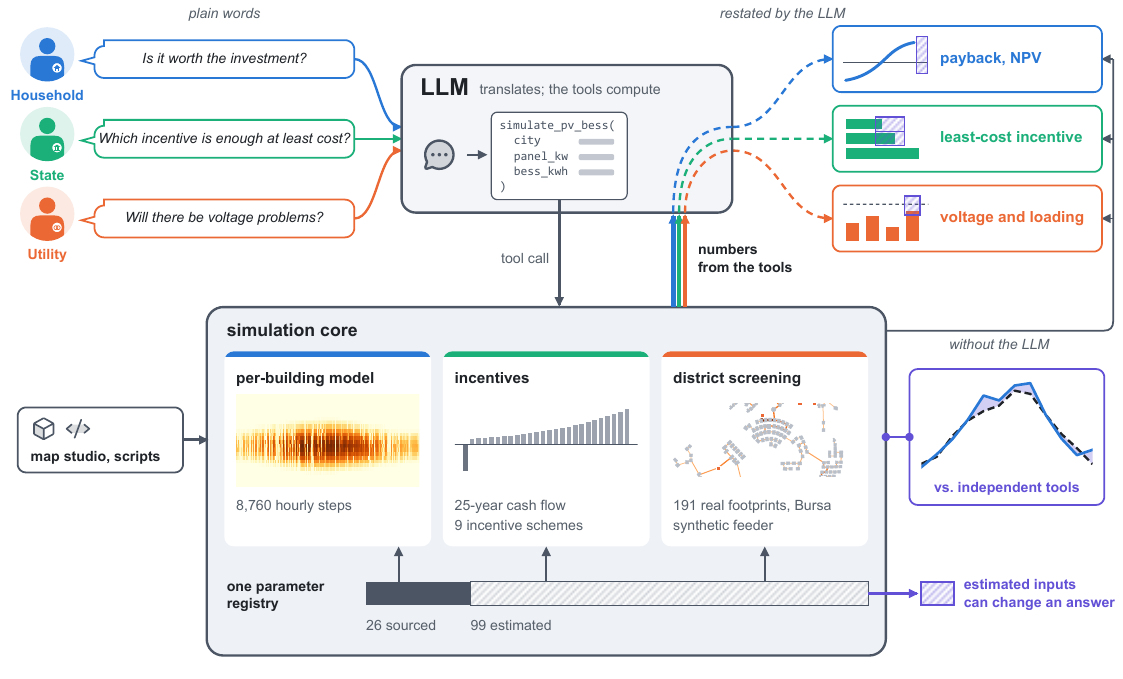}
\caption{Plain words above, expert tools below. The language model (LLM) is asked to turn the question into a tool call, not to compute. Tools in the simulation core compute the answer. The tools return the numbers, and the LLM restates them (dashed lines). The answer plots are schematic, not results. A guard withholds the model's text when it contains a number that matches no tool record; a matched number is not thereby shown to be the right one. An earlier version of the guard is evaluated in a companion study (\citealp{kaynak2026}). The map studio and scripts use the core without the LLM; only they run the Bursa district. Grey hatching marks estimated parameters. Violet marks the paper's checks: independent tools (RQ1) and, hatched on an answer, inputs that can change it (RQ3). No user study was run; Section~3 reports four scripted scenarios.}
\label{fig:1}
\end{figure}

The platform's parts agree with independent tools within stated gaps (RQ1). The district shortcuts err mainly with the size of the studio's default system (RQ2). The answers do not hold equally (RQ3). Under monthly netting and the supply tariff due from 2027, with prices corrected after the first runs (Section~6.1), the household's answer for PV alone held in all but one of 2,048 draws. With a battery, it changed in 31–41~\% of the draws, and the state's answer, that no incentive is needed, changed with it. On synthetic feeders, the utility's answer holds far from a limit and changes in about half of the draws just above it. In a district of 191 houses in Bursa, the default system times the number of houses gives 1.1 to 14.6 times the house-by-house state cost (RQ4).

Section~2 reviews related work, Sections~3 and 4 describe the platform and its models, and Sections~5–7 answer RQ1–RQ4. Section~8 lists the limits, Section~9 the software and data, and Section~10 concludes.

\section{Related work}
\textbf{How energy tools are checked.} \citet{boulaire2015} project how PV and electric-car uptake change the load on a real distribution network. They answer only the network planner's question, without economics, and check their modules against field records (regulator PV counts, measured feeder loads). MESSAGEix leans on openness because its long-term outputs cannot be checked against reality (\citealp{huppmann2019}); the parts of a rooftop model can be compared with other tools. Among the energy tools we reviewed, such a check is uncommon. \citet{brinkerink2022} compare one model with a more detailed tool, and \citet{su2020} theirs with observed data; \citet{chappin2017} report neither. Table~1 compares six related tools. URBANopt links building footprints, REopt sizing of PV and batteries, and OpenDSS feeder runs (\citealp{charan2021}; \citealp{urbanopt2026}). It gives the building and feeder answers, but it is a developer kit, not an end-user tool. In their own documentation, none ranks incentives by their cost to the state or reports how often an answer changes. Non-experts who explored a scenario tool understood less than those shown four fixed scenarios (\citealp{xexakis2019}). Whether our studio or chat helps understanding is untested.

\textbf{Sensitivity analysis for decisions.} Every such tool rests on uncertain inputs, and sensitivity analysis asks which of them matter. Global analyses of energy–economy models often rank inputs by their effect on a continuous output, as \citet{branger2015} do. Studies of PV incentives do the same for the net present value (\citealp{lyu2026}; \citealp{ardilacruz2027}). Neither study counts how often a decision itself changes. The two decision-oriented methods of Section~1 ask whether a decision changes, but for one decision between two options, and they bound or locate the change rather than estimate its share by sampling (\citealp{ravalico2010}; \citealp{guillaume2016}). No model we reviewed reports such shares for the household, state and utility answers together. We estimate them from one model and find the inputs behind each by Monte Carlo filtering (\citealp{spear1980}; \citealp{pianosi2016}).

\textbf{The cost to the state.} The state's answer has its own literature. Some studies price incentives on a few representative systems (\citealp{tantisattayakul2026}). Others add up the cost house by house or category by category, over tens of thousands of buildings to millions of households (\citealp{vaishnav2017}; \citealp{higgins2014}; \citealp{vimpari2019}). We test how far the shortcut lies from the house-by-house sum (RQ2). \citet{londo2020} compare Dutch instruments by their cost to the state at the same household payback. \citet{zakeri2021} compare UK instruments by household gain against regulator cost. Our state answer is narrower: it compares the instruments at fixed levels. \citet{higgins2014} found that the cheaper incentive depends on its level, which is why we vary incentive levels as inputs. Households discount future payments at about 15~\% a year, against about 3~\% for the state. An upfront subsidy therefore reached the same adoption at about half the budget of a production subsidy (\citealp{degroote2019}). The netting rule also changes the household's return (\citealp{dufolopez2015}), which our netting check in Section~7 measures.

\textbf{Synthetic feeders.} The utility's answer needs a feeder, and real feeder data were not available to us. \citet{krishnan2020} grade synthetic feeders, metric by metric, against the spread of real ones; we borrow that question, with European SimBench grids as the reference (\citealp{meinecke2020}). Two recent open tools also build low-voltage feeders from building footprints. \citet{weber2024} and \citet{reveron2025} route cables along streets; ours uses straight lines. On 128 real UK feeders, PV tended to break voltage limits first and electric cars to overload first (\citealp{navarro2016}), so we report both voltage and loading.

\textbf{Language models over simulators.} The chat interface raises a last question: how do language models reach simulators? In some tools the model writes the simulator's input (\citealp{dengler2023}; \citealp{becchi2025}). In others it fills a tool schema behind an interface that other applications also use (\citealp{mylonas2026}), or calls the tools of a Model Context Protocol (MCP) server (\citealp{liu2026}). PowerAgent proposes to connect a model to trusted power-system tools through MCP (\citealp{zhang2025}), and a building simulator has been exposed the same way (\citealp{li2025}). Our chat interface follows this pattern; its guard is designed and evaluated in a companion study (\citealp{kaynak2026}). None of the papers on these tools that we read compares the engine with independent tools or tests how often its answers change under its estimated inputs. We built the platform around seven requirements, which Section~3 sets out.

\begin{table}[tbp]
\centering
\caption{What six related tools do, from their own documentation, and what this paper adds. ``Not found'': not described in the sources read; ``no'': the source says the tool does not do it. Sources: SAM (\citealp{blair2018}; \citealp{freeman2014}), REopt (\citealp{cutler2017}; \citealp{mishra2022}), HOMER Pro and Grid (\citealp{homer2026}), dGen (\citealp{sigrin2016}), pylovo (\citealp{reveron2025}), URBANopt (\citealp{charan2021}; \citealp{urbanopt2026}).}
\label{tab:1}
\scriptsize
\setlength{\tabcolsep}{3pt}
\rowcolors{2}{black!6}{white}
\begin{tabularx}{\linewidth}{l>{\raggedright\arraybackslash}X>{\raggedright\arraybackslash}X>{\raggedright\arraybackslash}X>{\raggedright\arraybackslash}X>{\raggedright\arraybackslash}Xl}
\toprule
\textbf{Tool} & \textbf{Building NPV, PV and battery} & \textbf{Incentives by state cost} & \textbf{Feeder voltage and loading} & \textbf{Engine checked against} & \textbf{Input uncertainty} & \textbf{Licence} \\
\midrule
SAM & yes & not found & not found & measured data, other tools & parametric and stochastic runs & BSD-3 \\
REopt & yes, cost-optimal size & not found & no & not found & scenario batches & Apache/BSD \\
HOMER & yes & not found & no & not found & sensitivity grids over two inputs & commercial \\
dGen & partly, for sampled agents & partly: adoption, not cost & not found & calibrated to installation counts & Monte Carlo over agents & BSD-3 \\
pylovo & not found & not found & yes & reference grids, OpenDSS & one input at a time & MIT \\
URBANopt & yes, cost-optimal size (REopt) & not found & yes; feeders drawn or generated & not found & scenario runs & BSD-3 \\
\textbf{This paper} & yes & yes, nine instruments & yes, synthetic feeders & PVGIS, SAM, OpenDSS, national demand & 2,048 draws; how often each answer changes & MIT \\
\bottomrule
\end{tabularx}
\end{table}

\section{Architecture and clients}
Seven requirements shaped the platform. Table~2 names each, the mechanism behind it, its test and where the platform falls short; Section~8 returns to the shortfalls.

{\footnotesize
\begin{xltabular}{\linewidth}{>{\raggedright\arraybackslash}X>{\raggedright\arraybackslash}X>{\raggedright\arraybackslash}X>{\raggedright\arraybackslash}X}
\caption{Design requirements, the code that enforces them, and the main exceptions.}\label{tab:2}\\
\toprule
\textbf{Requirement} & \textbf{Mechanism} & \textbf{Test} & \textbf{Main exception} \\
\midrule
\endfirsthead
\toprule
\textbf{Requirement} & \textbf{Mechanism} & \textbf{Test} & \textbf{Main exception} \\
\midrule
\endhead
\bottomrule
\endlastfoot
R1 Numbers only from deterministic models & Tools call the simulators over REST; the guard withholds model text that contains a number matching no tool record (a matched number is not thereby right) & Guard, contract and orchestrator tests & A setting restores the older guard the companion study evaluates (Section~3.3); a rule-based reply after a model error is not checked \\
\arrayrulecolor{black!25}\midrule\arrayrulecolor{black}
R2 Every parameter sourced or marked estimated & A registry of 125 entries; a test binds each entry, and every constant of eight modules, to the code & Registry test & The demand-model constants, the cars' demand-factor constants, the cable type and the PV losses lie outside it; the grid emission factor (0.45) has a registry entry and three copies that do not read it \\
\arrayrulecolor{black!25}\midrule\arrayrulecolor{black}
R3 Fail loudly in the core & No clear-sky or synthetic-profile fallback; a missing input raises an error & No-silent-fallback test & A missing local footprint file returns a synthetic grid without a flag; the chat tools' defaults (Ankara, 5,000 kWh); in the code the companion study evaluated, the stored-system PV and battery tools fell back to clear sky and reported zero cycles (fixed in this release) \\
\arrayrulecolor{black!25}\midrule\arrayrulecolor{black}
R4 One core, two levels of detail & One feasibility function behind the studio, REST and tools; the district model reads the sizing constants & Defaults and contract tests & The policy tool has its own simplified cash flow; the two levels use different load models \\
\arrayrulecolor{black!25}\midrule\arrayrulecolor{black}
R5 Three decision views & Household, utility and state views, each on its own route & No end-to-end test & Policy numbers differ between the studio and the chat tool \\
\arrayrulecolor{black!25}\midrule\arrayrulecolor{black}
R6 Reproducibility & Hashed weather files, a replay mode for weather, seeds, frozen and tagged plans & Hash and replay tests & Hashes are checked by a test, not at run time; the result cache ignores the code version \\
\arrayrulecolor{black!25}\midrule\arrayrulecolor{black}
R7 Access without knowing the core & The studio sizes a system from a click on a house; the chat's tool schemas turn plain words into tool arguments & Four scripted scenarios test routing only (Section~3.4); no user study & Scenarios in English only, with the facts stated in the first turn \\
\end{xltabular}}

\subsection{Layers}

Fig.~2 shows the platform's layers and the software in each. The simulation core is Python. It holds the per-building model, the incentives and the district screening of Fig.~1, built on pvlib, demandlib and pandapower. One REST API, written with FastAPI and pydantic, is the only route by which the clients reach the core. The map studio, the network editor and the chat use it. The paper's analysis scripts import the core directly, so they run the same code as the clients. Weather comes from cached PVGIS typical years, each stored with its SHA-256 hash. Building footprints are read from a public PMTiles archive of the Google and Microsoft open building data. PostgreSQL keeps stored systems and networks, and Redis keeps chat sessions and a result cache. The platform's services run from one Docker Compose file; the Model Context Protocol (MCP) server runs in a container of its own.

\begin{figure}[tbp]
\centering
\includegraphics[width=\textwidth]{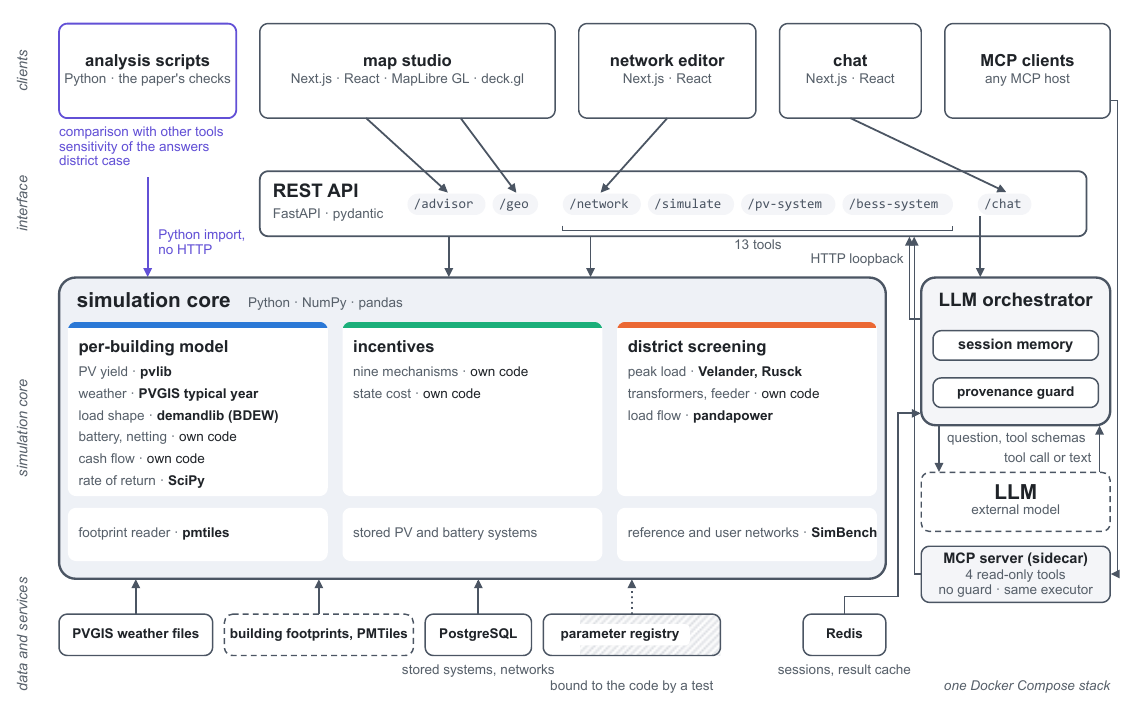}
\caption{Fig.~1 as built. The map studio, the network editor and the chat use one REST API; the chat's requests pass through the LLM orchestrator. MCP clients reach four of the 13 tools, read-only and without the guard, through a sidecar. The analysis scripts import the core directly (violet). Each line names the library or method doing the work. Hatching: the estimated part of the parameter registry (Fig.~1's ``one parameter registry''). Dashed: outside the stack; dotted: bound to the code by a test, not read at run time.}
\label{fig:2}
\end{figure}

\subsection{Clients without the language model}

Fig.~2 shows two clients that skip the LLM: the map studio and the network editor. Fig.~3 shows their screens. The studio opens on a 3D map with the three views of Fig.~1. In the household view, a click on a house reads its footprint, lays out panels on the roof and sizes PV and a battery from a monthly bill (Fig.~3a). The state view compares the nine incentives for one default system, scaled to a number of systems (Fig.~3b). The utility view builds the synthetic feeder for the buildings in view and runs the load flow; transformers and cables can be moved by hand (Fig.~3c). The studio's defaults are 40~\% PV and 3~\% electric cars; Fig.~3c raises the car share to 30~\%, and the frozen runs of Sections~5–7 use none. The studio takes its weather at the map centre, the runs at the Bursa site point (Section~7).

\begin{figure}[tbp]
\centering
\includegraphics[width=\textwidth]{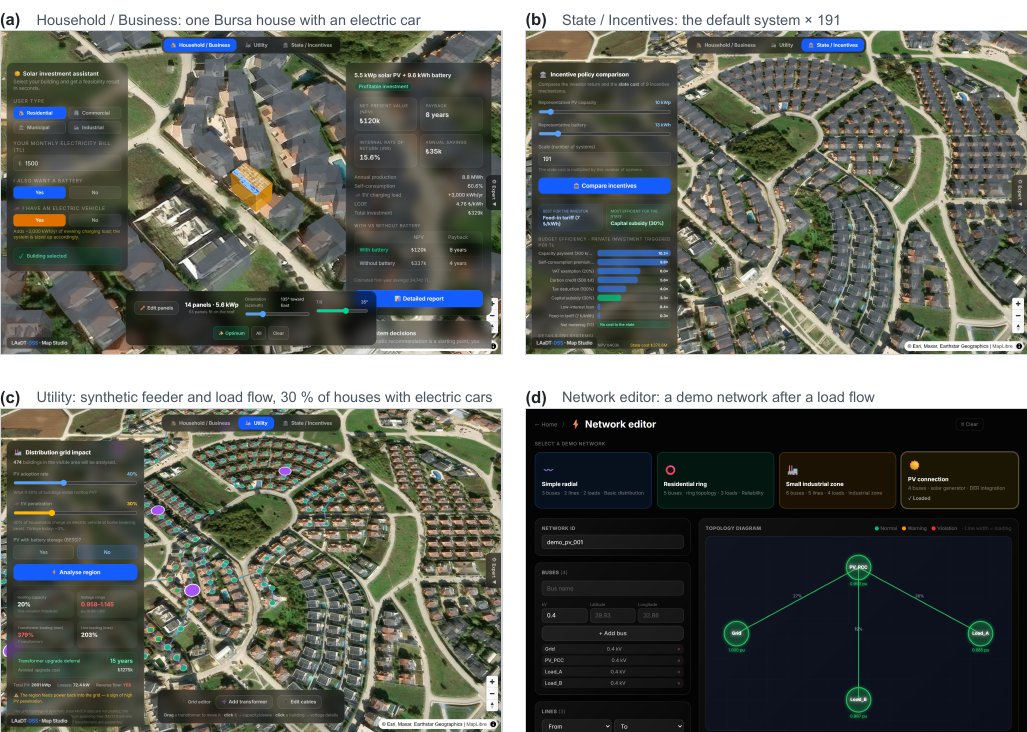}
\caption{The clients of Fig.~2 that skip the LLM, in English mode (default: Turkish). (a)–(c) The map studio over the Bursa district of Section~7. (a) A clicked house with one electric car, with PV and a battery sized from a monthly bill and costed. (b) The nine incentives for the studio's default system (10 kWp), scaled to 191 systems. (c) The district screening on all 474 footprints in view, with 30~\% electric cars; Section~7 screens only its 191 houses. (d) The network editor on a demonstration network. Feeders and numbers on screen are the studio's outputs, not this paper's results. Basemap: © Esri, Maxar, Earthstar Geographics.}
\label{fig:3}
\end{figure}

The network editor builds a network bus by bus and runs a load flow on it (Fig.~3d).

\subsection{The language-model loop}

Fig.~4 follows one chat turn from a plain question to the core. The orchestrator sends the question, the chat so far and the schemas of 13 tools to the model, with a system prompt for the user's role (Turkish by default; the scenario runs set English). Four tools run simulations (PV and battery, grid impact, policy and Monte Carlo); nine create, change or run stored systems and networks (Fig.~4b). The model answers with a tool call or with text. Each tool call goes to the REST API (Fig.~4a, step 4), and its result returns to the model, for at most four tool-call turns. If the model call fails, a rule-based responder answers instead, and the response is flagged.  Before the reply is shown, a provenance guard matches the numbers in the model's text against the tool records (Section~3.4). The server's chat and the scenario runs of Section~3.4 use a successor of the orchestrator, with fixes and later versions of the guard; the MCP server shares its tool executor. A configuration setting restores the frozen version evaluated in the companion study (\citealp{kaynak2026}); that study did not evaluate the later versions.

\begin{figure}[tbp]
\centering
\includegraphics[width=\textwidth]{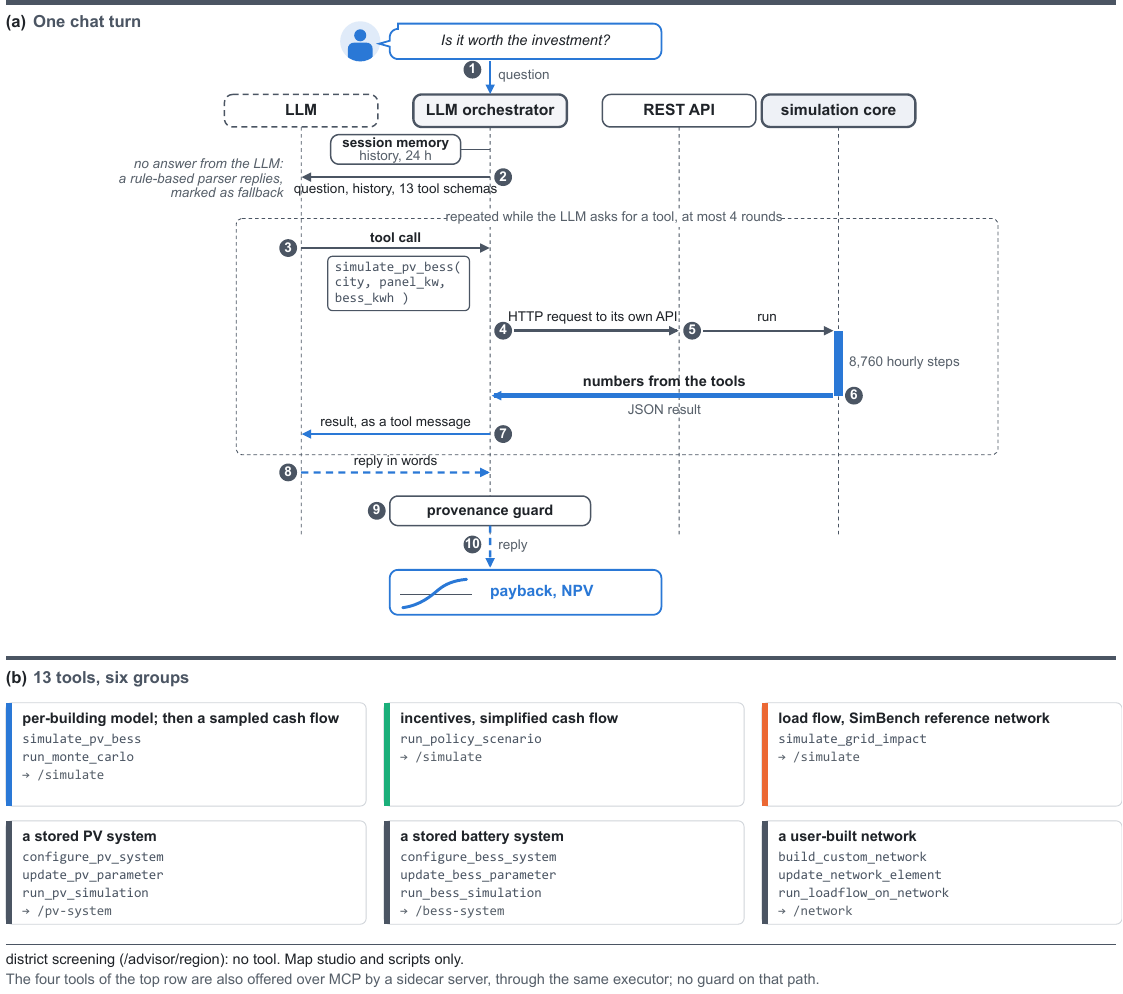}
\caption{The LLM box of Fig.~1, one turn. (a) The LLM is asked to turn the question into a tool call, not to compute. The orchestrator sends each call to the REST API, for at most four rounds. A guard withholds the model's text when it contains a number that matches no tool record; a matched number is not thereby shown to be the right one. An earlier version of the guard is evaluated in a companion study (\citealp{kaynak2026}). Dashed: the external model and its text. (b) The 13 tools; bar colours: the Fig.~1 answer each group serves (grey: none). Grid questions run on a SimBench reference network, not on the district screening. Last line: the MCP path, without a guard.}
\label{fig:4}
\end{figure}

The chat is not the only way for a language model to reach the core. A small MCP server lets an MCP host reach it as well (Fig.~2, right; Fig.~4b, last line). Other energy tools also expose their functions over MCP (\citealp{li2025}; \citealp{liu2026}). Our server offers four of the 13 tools, those that change no stored system or network (Fig.~4b, top row). Each call goes through the same code that executes tool calls in the scenario runs (Section~3.4). An end-to-end test opens an MCP session, runs one PV case and gets the same numbers as the executor called directly. This path has no guard. The client's own model writes the reply, and the platform never sees that text, so the guard results of the companion study do not apply to MCP clients. Each result also carries provenance records: one per numeric input and up to 300 for the result's numbers, each naming its source and the call's scalar inputs. They show where a number came from; they do not check the reply. The tools that change stored data stay closed to MCP: MCP enforces no least privilege, and many users let hosts run tools without approval (\citealp{zhao2026}). The scenarios that follow call the orchestrator directly, not through MCP.

\subsection{Use-case scenarios}

The interface above is meant to let a person who knows nothing about the simulation core reach it in plain words. Does a plain request reach the right tool with the right arguments? We ran no user study. Instead, we wrote four scenarios for the roles the platform serves. The roles are a homeowner, a grid planner and two policy makers: a municipal energy manager and a ministry expert. The scenarios demonstrate the interface; they do not evaluate the language model, which is the subject of the companion study (\citealp{kaynak2026}).

Each scenario has two user turns, in English, frozen with a git tag before any run. The first turn states the facts a tool needs: city, system size and annual consumption. The second asks a follow-up question, such as ``what if I add a battery?''. The model, \texttt{deepseek-v4.\allowbreak{}1-flash}, ran each scenario three times at temperature 0.3 with fixed seeds. We report the first of the three; the transcripts of all runs are in the archive (Section~9). The platform's provenance guard was on, as it is by default; its design is the companion study's. It withholds model text that contains a number matching no tool record and shows a summary built from the tool records instead, the verbatim-provenance summary. The companion study evaluates the guard's frozen form, not the later versions also used here.

The runs exposed five problems that no unit test, verification run (Section~5) or sensitivity run (Section~6) had touched, and work on the companion study found a sixth; Table~S1 in the Supplement lists all six. Three problems sat in the guard code the platform inherited (problems 1, 2 and 6); all three are also described in the companion study (in preparation). First, the verbatim-provenance summary reads eight fixed fields at the top level of the last tool result; it is built from records and is not itself re-checked. Second, the matcher compares magnitudes and does not read a numeral's sign, so a correctly quoted negative value is unverifiable, and a sign-inverted value can match a positive record. Third, a numeral with one to three integer digits and four or more decimals is split, leaving a stray digit that can be flagged. The other three problems sat in the platform's own tools. Two of them changed a decision-relevant number: a fivefold error in first-year savings (problem 3) and a loan priced at zero (problem 5).

We fixed the problems in three batches, all in new modules, and left the frozen guard and orchestrator unchanged, byte for byte, so that the companion study can be reproduced. The server's chat now runs the new modules by default; a configuration setting restores the frozen orchestrator. After each batch we reran the frozen user turns, and each fix was committed with its reason before its rerun. With the first round, this gives four rounds of runs.

Fixes 1 and 2 sit in a new guard module, which we call v5. Fix 6 changes how the guard reads numbers, so it touches what the companion study measured. We therefore keep it apart, in a second new module, v6.

What does each role receive in round 4, at the prices the tools then used (Section~6.1)? In every first run of round 4, the model called the tool that answers the role's question, with the arguments taken from the user's words. The homeowner learns that PV alone pays back in about six years, and that a battery turns the NPV negative. The grid planner learns that the SimBench reference network (Section~4) breaks its voltage limits even without PV. The share of PV it can host before a limit is crossed is therefore 0~\%, an illustrative figure set by that network (Section~8). The municipal manager receives an NPV band that straddles zero. The ministry expert finds that the loan, which looked free before fix 5, is the most expensive incentive per house. At the corrected 2027 prices, the homeowner's battery answer turns to ``invest'' at the three sites of Section~6.1. The released studio and chat keep the older prices of Section~6.1; only the analysis scripts apply the newer ones.

These runs do not show that the model's explanations are accurate, nor that real users can work with the interface. They say nothing about other models, or about Turkish, the language of the platform's intended users. What they do show is narrower. A few scripted conversations, run end to end, found errors that component tests had missed. Problem 4, a policy-tool schema that did not show the model what a valid incentive package looks like, marks the limit of the contract test described next. That test checks that every argument reaches its REST route. It does not check that a schema tells the model what a valid argument looks like.

\subsection{Contract test}

The contract test checks the path between the parts. It sends each property of each tool schema alone, with a value that differs from the defaults, and records the HTTP calls the orchestrator makes. All 62 properties of the 13 tools reach their REST route. Three update tools also remove the identifier from the caller's argument dictionary. The test reports this rather than fixing it, because the frozen orchestrator is kept unchanged so that the companion study can be reproduced. The test ran against that frozen orchestrator. The successor the server runs inherits the same routes and adds one property, the Monte Carlo tool's stakeholder type, which its own test covers. A second test compares each request type of the frontend with the pydantic model of its route. One field, a panel type that the PV-and-battery route does not use, is dropped silently. These tests cover the paths between the parts, not what runs inside the core.

\section{Models and parameters}
Fig.~A.1 gives the two chains that compute the numbers, with the main equation of each step.  Appendix~A gives every equation as the code computes it.

\textbf{Per building.} The first chain answers the household and state questions of Fig.~1 for one building (Fig.~A.1a). PV output comes from pvlib for each hour of a typical year from PVGIS at the site (E1; \citealp{anderson2023}; \citealp{huld2012}). The chain uses Hay–Davies transposition (\citealp{hay1980}), the SAPM cell temperature (\citealp{king2004}) and the PVWatts models for the module and the inverter (\citealp{dobos2014}). The load takes the shape of the German standard household or commercial profile (BDEW; \citealp{meier1999}), scaled to the annual consumption (E2). A battery, if present, charges from surplus PV and discharges to cover the load, within fixed power and charge limits (E3). Imports and exports are netted per month for households and per hour for others, as the regulation on unlicensed generation sets (E4; \citealp{epdk2026a}). From year 11, households move to hourly netting, with the excess paid at most the market price (\citealp{presidency2026}); the model keeps monthly netting for all 25 years (Section~8). Below 3,000 kWh a year a household pays the national two-tier tariff of April 2026 (\citealp{epdk2026tariff}): each month, consumption up to 8 kWh a day costs 3.238 TL/kWh with taxes and the rest 4.857 TL/kWh. We apply the tiers to the netted monthly imports [post-outcome]; this tag marks what we added after seeing results. From 2027, households that use 3,000 kWh a year or more pay the last-resort supply tariff, whose energy price follows the market (\citealp{epdk2026b}); over the twelve months to September 2026 (\citealp{epias2026b}) it comes to 6.096 TL/kWh with taxes. Sections~6 and 7 use these prices where they apply. Net exports earn the high-tier active-energy price, 1.8958 TL/kWh, our reading of article 28(11) of the regulation. The battery's ageing combines a calendar term and a cycling term from a rainflow count (after \citealp{schmalstieg2014}; \citealp{ecker2012}), and it sets only the year in which the battery is replaced (E5). The savings then run through a 25-year cash flow in nominal lira with escalation, upkeep, replacements and degradation, and give the net present value (NPV) and the payback year (E6, E7). The incentives of the state view change this cash flow and carry their own state cost (E8, E8a).

\textbf{District screening.} The second chain screens a feeder for many buildings at once (Fig.~A.1b).  Each building's annual energy follows from its footprint area: 2,800 kWh for a 120 m² dwelling, clipped to between half and three times that (E9). Peak loads follow from the Velander formula (\citealp{dickert2010}), and a coincidence factor reduces their sum as the number of houses grows (E10; \citealp{furst2020}). The platform places transformers by k-means clustering, sizes each for the coincident peak, and links the houses by a minimum spanning tree of one cable type (E11). A load flow in pandapower (\citealp{thurner2018}) then runs at two operating points: the evening at full load without PV, and noon at low load with PV on a share of the houses (E12, E13). The two are kept apart, because PV at noon and load in the evening would cancel in one snapshot.

\textbf{Origins.} The tariff, the netting rule and the irradiance are Turkish. The household and commercial load shapes are German. The reference grids, used for comparison and by the language model's grid tool, are German SimBench networks (\citealp{meinecke2020}). Nothing in the two chains was fitted to Turkish measurements.

\textbf{Parameter registry.} The numbers the chains use are listed in one registry, with value, unit, status and place in the code. It holds 125 entries. Of these, 26 cite a source, 21 of them in the tariff. The other 99 are estimates: modelling choices without a cited source, such as the battery's ageing constants, the cost per kWp or the incentive levels. A test binds every entry to the code. For eight modules the test also checks that no constant is missing. The registry is not read at run time. Its coverage has gaps: the demand-model constants of E10 and the cable type of E11 have no entry, and Fig.~A.1 draws them dotted. In the screening chain, only the 2,800 kWh per dwelling has a stated derivation (national residential consumption divided by the number of households); its inputs were not checked against the published tables.

An estimate is not a wrong value. It is a value for which we could cite no source. Section~5 gives the comparisons; Section~6 moves the estimates.

\section{Verification}
Fig.~A.1 names, for each step, what it was compared with. Do they agree (RQ1)? Fig.~5 compares the PV yield with PVGIS and SAM, and the load shape with the measured national demand of 2023. Table~3 adds the load flow and the cash flow. Two further checks follow: whether the synthetic feeders look like real ones, and how far the district shortcuts lie from the per-building model (RQ2). All settings, metrics and building sets were fixed in a plan committed before any comparison ran. The building sets were later de-duplicated, and results on them come from reruns with the frozen rules [post-outcome].

\begin{figure}[tbp]
\centering
\includegraphics[width=\textwidth]{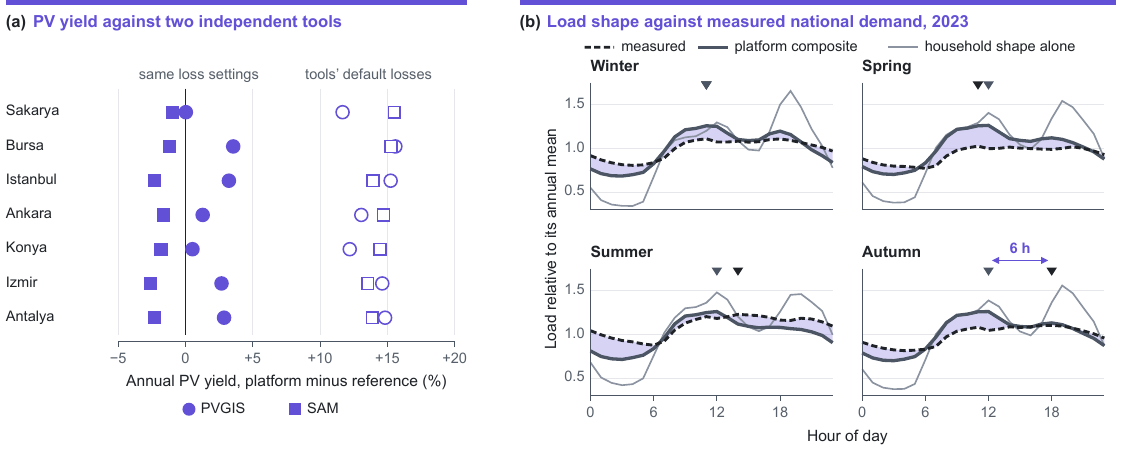}
\caption{The violet tile of Fig.~1 (``vs. independent tools''). (a) Annual PV yield at seven sites, platform minus reference. Filled marks: the same loss settings in both (the platform's); open marks: each tool's defaults. (b) Mean daily load by season, relative to its annual mean. Dashed: measured national demand; solid: the platform's composite of household, commercial and flat shapes; grey: the household shape alone. Violet fill: the difference. Triangles: the daily peaks (dark: measured; grey: platform); ``6 h'': the gap between them in autumn.}
\label{fig:5}
\end{figure}

\textbf{PV yield.} We compared a 1 kWp system, tilted 30° to the south, at seven Turkish sites. With the same loss settings as the platform, the annual yield lies within ±3.6~\% of both PVGIS PVcalc and SAM's PVWatts v8 model (PySAM 7.1.1; \citealp{blair2018}, for SAM; Fig.~5a, filled marks). With each tool's default losses, the platform is 12–16~\% higher (open marks). It models no soiling, wiring or mismatch losses, so its yield is optimistic.

\textbf{Load shape.} The measured series is national, while the platform gives each building one shape. For the comparison we therefore built a composite from three shapes. The household and commercial shapes take 26~\% each and a flat share 48~\%, as in national consumption in 2023 (\citealp{epdk2024}). Against measured national demand (\citealp{epias2026a}), its hourly correlation over the year is 0.65.  The mean daily shapes agree by season (r = 0.82–0.93), and their peak hour is off by 3.0 h over the twelve months. In winter both peak at 11 h (the measured curve is almost as high at 18 h). In autumn the measured peak comes at 18 h and the composite's at 12 h, six hours early. The normalised RMSE (0.155) is no better than a flat line's (0.160) [post-outcome]. The flat share dominates it, so we rely on the correlations and the peak hours, which do not depend on that share.

\begin{table}[tbp]
\centering
\caption{The load flow and the cash flow against independent tools.}
\label{tab:3}
\footnotesize
\rowcolors{2}{black!6}{white}
\begin{tabularx}{\linewidth}{l>{\raggedright\arraybackslash}X>{\raggedright\arraybackslash}X>{\raggedright\arraybackslash}X}
\toprule
\textbf{Component} & \textbf{Reference} & \textbf{Case} & \textbf{Result} \\
\midrule
Load flow & OpenDSS (\citealp{dugan2011}), same networks, positive sequence & Three synthetic feeders (116–260 buses), evening and noon & Voltage within 0.00021 pu; loading within 0.08 percentage points \\
Load flow & OpenDSS & SimBench 1-LV-rural1 (15 buses) & Voltage within 0.00013 pu; loading within 0.005 percentage points \\
Cash flow & SAM, same first-year inputs & PV only, 5 kWp, 4,000 kWh/yr, Sakarya and Ankara, hourly netting, low-tier export price (Section~6.1) & First-year savings +1.3 and +0.3~\%; NPV −9.8 and −11.5~\% \\
\bottomrule
\end{tabularx}
\end{table}

\textbf{Load flow.} This tests the solver and our network export, not the feeder model. The two solvers agree to below 0.001 pu at both operating points (Table~3). At 40~\% adoption, the synthetic feeders carry PV on transformers sized for the evening load without it. At noon, transformer loading then reaches 448–936~\% and line loading 238–501~\% [post-outcome]. Section~6 therefore tests noon at lower PV levels.

\textbf{Feeder plausibility.} We compared the 14 synthetic transformer networks with six SimBench low-voltage grids. Within SimBench's range lie 12 of them on customers per transformer, all 14 on line length per customer and 9 on feeder depth. On the minimum evening voltage, the metric fixed in advance, 10 of 14 lie within range. Measured as the voltage drop from the slack bus, 12 of 14 do [post-outcome]. The two outside are feeders with one and nine customers, which the automatic transformer placement produced.

\textbf{Cash flow.} The first-year savings agree with SAM within 1.3~\%, but the net present value is 10–12~\% lower. A check fixed before the run fed SAM's first-year savings into the platform's rules and missed SAM's NPV by more than 2~\%. Traced after the run, the gap is the degradation rule: escalation, upkeep and discounting agree exactly. The platform reduces the whole annual savings by 0.5~\% a year. SAM reduces the energy and bills it again. The energy lost is mostly cheap export, so SAM's savings fall by about 0.2~\% a year. This rule alone puts the platform's NPV 13–18~\% below SAM's. Slightly higher first-year savings recover part of it, which leaves the 10–12~\%. This comparison used hourly netting; the monthly netting behind the household answers of Section~6 was not compared with another tool.

\textbf{Screening against the per-building model (RQ2).} The platform's screening shortcuts assume a fixed yield of 1,400 kWh per kWp. At the Sakarya site the per-building model gives 1,418.5, so screening over per-building is 0.987. Across the seven sites the ratio runs from 0.74 (Antalya) to 0.99, so Sakarya is the most favourable case [post-outcome]. The state view of Fig.~1 multiplies one default system by the number of buildings; we used its PV-only form (10 kWp, 4,000 kWh a year). On 498 buildings in three Sakarya test areas (after removing footprints that the provider returned twice along tile edges), this understated the state cost and the avoided CO\textsubscript{2} of the house-by-house sum by 58–87~\% (median over the nine incentives). These are large roofs, sized by area and mostly exporting. The cause is the default size, not the multiplication. A representative with each area's mean size and consumption reproduces the state-cost and CO\textsubscript{2} sums within 0.81~\% [post-outcome].

So the core agrees with independent tools where both model the same thing. Three gaps have a traced cause: losses not modelled, the degradation rule and the default system size. The fourth, the missed autumn evening peak, does not. All these comparisons used the registry values.

\section{How stable are the answers?}
\subsection{Do the answers survive the estimates?}

The comparisons of Section~5 all used the registry values, and most of these values are estimates (Fig.~1, grey hatching). We moved 29 inputs together, inside their declared ranges, over 2,048 independent draws (Table~4).

{\footnotesize
\begin{xltabular}{\linewidth}{l>{\raggedright\arraybackslash}Xlll}
\caption{The inputs varied in Section~6: registry value and range. Household and state answers: the 29 rows above the feeder group (16 model estimates, 3 structural factors, 10 incentive levels). Feeder answer: the 7 feeder rows. The 10 incentive levels are design choices rather than estimates. Household and state draws are independent and uniform within each range; the feeder rows use the quasi-random A and B rows of the feeder's own Sobol sample (Section~6.3). The Sobol cross-check of Section~6.1 uses a base sample of 2,048 and a dummy input (22,528 rows per case). Held fixed (sourced): the import price (the last-resort supply tariff for the seven cases, Section~4) and the export price, 1.8958 TL/kWh.}\label{tab:4}\\
\toprule
\textbf{Group} & \textbf{Input} & \textbf{Unit} & \textbf{Registry value} & \textbf{Range} \\
\midrule
\endfirsthead
\toprule
\textbf{Group} & \textbf{Input} & \textbf{Unit} & \textbf{Registry value} & \textbf{Range} \\
\midrule
\endhead
\bottomrule
\endlastfoot
Model estimate & PV cost & TL/kWp & 25,000 & 12,500–37,500 \\
\arrayrulecolor{black!25}\midrule\arrayrulecolor{black}
Model estimate & Battery cost & TL/kWh & 20,000 & 10,000–30,000 \\
\arrayrulecolor{black!25}\midrule\arrayrulecolor{black}
Model estimate & Tariff growth & 1/yr & 0.08 & 0.04–0.12 \\
\arrayrulecolor{black!25}\midrule\arrayrulecolor{black}
Model estimate & Discount rate & 1/yr & 0.12 & 0.06–0.18 \\
\arrayrulecolor{black!25}\midrule\arrayrulecolor{black}
Model estimate & Lifetime & yr & 25 & 20–30 \\
\arrayrulecolor{black!25}\midrule\arrayrulecolor{black}
Model estimate & PV upkeep & of cost/yr & 0.01 & 0.005–0.015 \\
\arrayrulecolor{black!25}\midrule\arrayrulecolor{black}
Model estimate & Inverter replacement & of cost & 0.1 & 0.05–0.15 \\
\arrayrulecolor{black!25}\midrule\arrayrulecolor{black}
Model estimate & Battery replacement & of first cost & 1 & 0.5–1 \\
\arrayrulecolor{black!25}\midrule\arrayrulecolor{black}
Model estimate & Battery end-of-life loss & – & 0.2 & 0.15–0.3 \\
\arrayrulecolor{black!25}\midrule\arrayrulecolor{black}
Model estimate & Battery calendar ageing k & 10⁻³ & 1.05 & 0.525–1.575 \\
\arrayrulecolor{black!25}\midrule\arrayrulecolor{black}
Model estimate & Battery cycle ageing k & 10⁻⁵ & 5.5 & 2.75–8.25 \\
\arrayrulecolor{black!25}\midrule\arrayrulecolor{black}
Model estimate & Battery ageing exponent & – & 1.2 & 0.6–1.8 \\
\arrayrulecolor{black!25}\midrule\arrayrulecolor{black}
Model estimate & Battery round-trip efficiency & – & 0.9 & 0.8–0.95 \\
\arrayrulecolor{black!25}\midrule\arrayrulecolor{black}
Model estimate & Battery minimum charge & – & 0.1 & 0.05–0.2 \\
\arrayrulecolor{black!25}\midrule\arrayrulecolor{black}
Model estimate & Battery maximum charge & – & 1 & 0.9–1 \\
\arrayrulecolor{black!25}\midrule\arrayrulecolor{black}
Model estimate & Battery charge rate & C & 0.5 & 0.25–1 \\
\arrayrulecolor{black!25}\midrule\arrayrulecolor{black}
Structural & PV yield factor & – & 1 & 0.84–1 \\
\arrayrulecolor{black!25}\midrule\arrayrulecolor{black}
Structural & Cost escalation & 1/yr & 0.08 & 0–0.12 \\
\arrayrulecolor{black!25}\midrule\arrayrulecolor{black}
Structural & Savings degradation & 1/yr & 0.005 & 0.002–0.005 \\
\arrayrulecolor{black!25}\midrule\arrayrulecolor{black}
Incentive level & Capital subsidy & \% of cost & 30 & 15–45 \\
\arrayrulecolor{black!25}\midrule\arrayrulecolor{black}
Incentive level & Feed-in tariff & TL/kWh & 7 & 3.5–10.5 \\
\arrayrulecolor{black!25}\midrule\arrayrulecolor{black}
Incentive level & Loan share & \% of cost & 80 & 40–100 \\
\arrayrulecolor{black!25}\midrule\arrayrulecolor{black}
Incentive level & Loan rate & \%/yr & 15 & 7.5–22.5 \\
\arrayrulecolor{black!25}\midrule\arrayrulecolor{black}
Incentive level & Loan term & yr & 10 & 5–15 \\
\arrayrulecolor{black!25}\midrule\arrayrulecolor{black}
Incentive level & Market rate (state's loan cost) & \%/yr & 45 & 22.5–67.5 \\
\arrayrulecolor{black!25}\midrule\arrayrulecolor{black}
Incentive level & Self-consumption premium & TL/kWh & 0.5 & 0.25–0.75 \\
\arrayrulecolor{black!25}\midrule\arrayrulecolor{black}
Incentive level & Carbon credit & TL/t & 500 & 250–750 \\
\arrayrulecolor{black!25}\midrule\arrayrulecolor{black}
Incentive level & Tax deduction & \% of cost & 100 & 50–100 \\
\arrayrulecolor{black!25}\midrule\arrayrulecolor{black}
Incentive level & Capacity payment & TL/kW/yr & 200 & 100–300 \\
\arrayrulecolor{black!25}\midrule\arrayrulecolor{black}
Feeder & PV power per roof area & kWp/m² & 0.14 & 0.07–0.2 \\
\arrayrulecolor{black!25}\midrule\arrayrulecolor{black}
Feeder & PV noon factor & – & 0.8 & 0.6–1 \\
\arrayrulecolor{black!25}\midrule\arrayrulecolor{black}
Feeder & Daytime load factor & – & 0.4 & 0.2–0.6 \\
\arrayrulecolor{black!25}\midrule\arrayrulecolor{black}
Feeder & Transformer sizing factor & – & 1.25 & 1–1.5 \\
\arrayrulecolor{black!25}\midrule\arrayrulecolor{black}
Feeder & Buildings per transformer & – & 40 & 20–60 \\
\arrayrulecolor{black!25}\midrule\arrayrulecolor{black}
Feeder & Maximum transformers & – & 8 & 4–12 \\
\arrayrulecolor{black!25}\midrule\arrayrulecolor{black}
Feeder & Typical dwelling area & m² & 120 & 60–180 \\
\end{xltabular}}

An answer changes in a draw when it differs from the answer with the registry values. Fig.~6 (a, b) shows the share of draws in which the answer changed. This share answers RQ3. It is not a probability, since the ranges come from a rule, not from data; we give it with a 95~\% Wilson interval (\citealp{wilson1927}).

\begin{figure}[tbp]
\centering
\includegraphics[width=\textwidth]{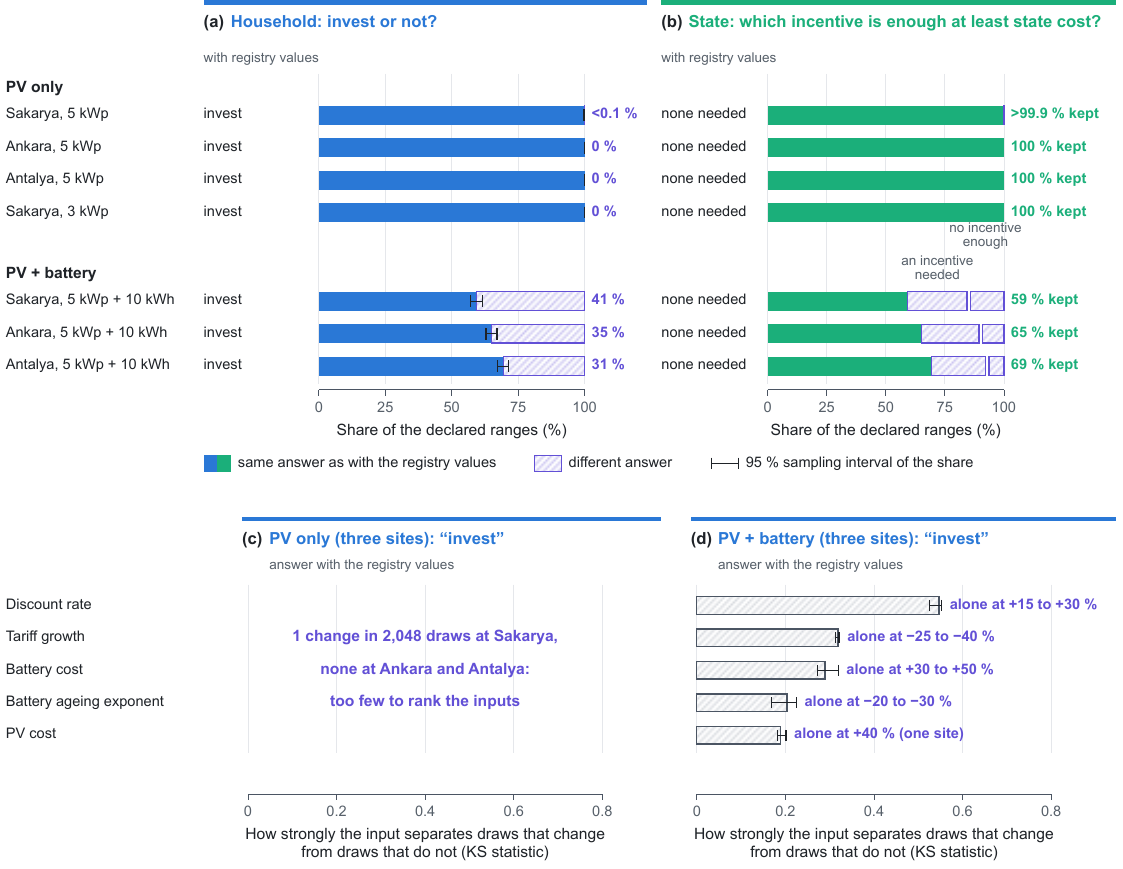}
\caption{The Household and State answer cards of Fig.~1 with data. Solid: the same answer as with the registry values. Violet hatching: a different answer. Whisker in (a): 95~\% sampling interval. (a) Invest or not (NPV above zero). (b) The incentive that is enough at least state cost; the changed share is split into draws that need an incentive and draws in which none is enough. (c, d) The estimated inputs that separate draws in which the answer changed from the others (KS statistic); PV alone (c) changed too rarely to rank them. In (d), the bar is the median of three sites and the whisker their range. Violet: the nearest shift of that input alone that changes the answer. Prices from 2027, corrected after the outcome (Section~6.1). All inputs shown separate the draws at KS p < 0.01, a threshold set after the outcome; with p ≤ 9.3 × 10⁻¹³, they also pass a Bonferroni threshold of 0.01/29.}
\label{fig:6}
\end{figure}

The household and state answers are tested here, the feeder answer in 6.3. We used seven cases, each with 4,000 kWh a year: 5 kWp of PV alone and the same with a 10 kWh battery, each at Sakarya, Ankara and Antalya, and 3 kWp of PV alone at Sakarya. The three sites share the same draws. The household answer is ``invest'' when the NPV is positive. The state answer is the incentive that makes it positive at the least state cost (Appendix~A, E8). Most ranges run from half to one and a half times the registry value. Five run only downward from it: the PV yield factor (0.84–1.0), the battery replacement cost, the battery's maximum charge, the savings degradation and the tax deduction. All of this was frozen and tagged before the first run. The inputs behind a change are found by Monte Carlo filtering (\citealp{spear1980}; \citealp{pianosi2016}). Sobol indices (\citealp{saltelli2010}) on eight inputs selected by Morris screening (\citealp{morris1991}), computed with SALib (\citealp{herman2017}), serve as a cross-check. Single-input break-even points, the smallest shift of one input that changes the answer, give a local view (\citealp{guillaume2016}).

The frozen runs priced, and the released studio still prices, exports at the low tier, 0.4941 TL/kWh, and imports at the national tariff, which these households leave in 2027 (Section~4). With those prices, the battery answer was ``do not invest'' at all three sites. We reran the frozen design with the corrected prices [post-outcome]. Table~S2 gives the frozen results, a rerun under the 2026 tariff and the 2027 rerun reported here, with the feed-in tariff paid on billed exports only (Section~8). A probe on the Bursa households had shown the direction of the export-price change before that rerun was planned, and the setting reported here was chosen after both reruns.

With the registry values and the monthly netting of households, PV alone is worth investing in at all three sites, and no incentive is needed. The answer held in all but one of the 2,048 draws for 5 kWp, and in every draw for 3 kWp (Fig.~6a). With a battery, the answer is also ``invest'', and it changes in 30.7–40.7~\% of the draws.

The state answer follows. With the registry values no incentive is needed, so the state answer changes exactly when the household answer does. When the battery household stops investing, no incentive is enough in 7.8–15.6~\% of all draws, and seven or eight instruments share the rest (Fig.~6b). The low-interest loan almost never wins: at the registry rates (loan 15~\%, discount 12~\%) it lowers the household's NPV, and it wins in one draw at Sakarya and none elsewhere. The incentive levels are inputs here, and three of the instruments are upfront transfers that differ only in level. Which of them is cheapest therefore shows where the household's shortfall falls between the studio's levels: the state answer is the tool's output at those levels, not a ranking of instruments.

\subsection{Which inputs change the answers?}

For each input we compared the draws in which the answer changed with those in which it did not, using the Kolmogorov–Smirnov (KS) statistic (Fig.~6c, d). PV alone changes too rarely to rank its inputs.

For PV with a battery, the discount rate leads at all three sites (KS 0.53–0.55), followed by the tariff growth (0.31–0.32), the battery cost (0.27–0.32), then the battery ageing exponent (0.17–0.22) and the PV cost (0.18–0.20), in an order that varies by site (Fig.~6d). The first four have a break-even point inside their range at every site; the PV cost only at Sakarya (+40~\%). For the discount rate it lies at +15 to +30~\%, from 12 to 13.8–15.6~\%. The tariff growth needs −25 to −40~\%, the battery cost +30 to +50~\% and the ageing exponent −20 to −30~\%. Sobol indices give the same leader. Its total index is 0.53–0.58 for the invest-or-not answer itself and 0.45–0.47 for the NPV. These indices converged; those for the choice of incentive did not. With the first three kept to the inner half of their ranges, the battery answer still changes in 23.0–37.5~\% of the draws [post-outcome]. Its fragility does not come from the edges of the ranges alone.

\subsection{The feeder answer}

Fig.~7 orders nine feeder cases by their highest loading with the registry values.

\begin{figure}[tbp]
\centering
\includegraphics[width=\textwidth]{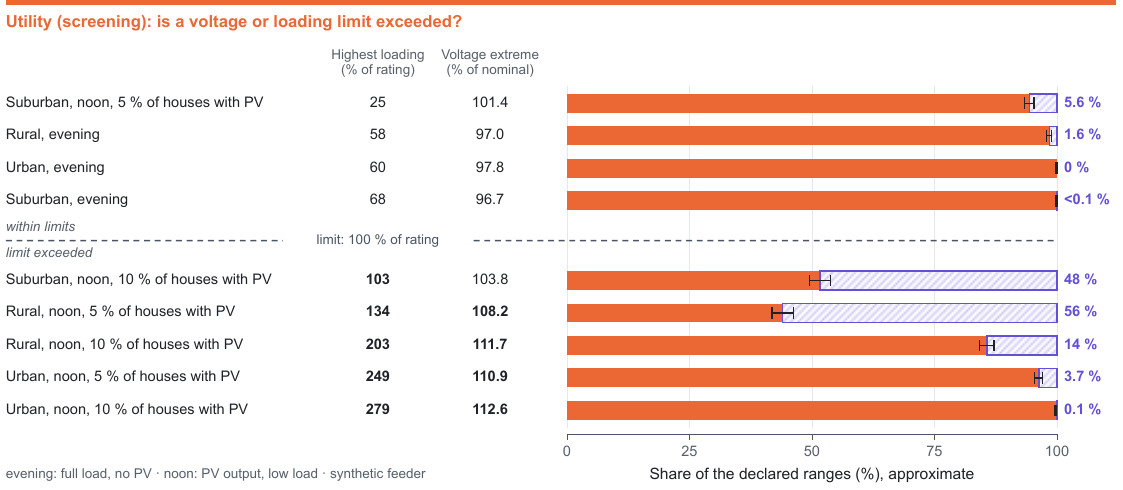}
\caption{The Utility card of Fig.~1 (``voltage and loading'') with data. Nine screening cases on synthetic feeders in three Sakarya test areas (urban, suburban, rural), ordered by highest loading with the registry values. Solid: the same answer; violet hatching: a different answer (seven estimated inputs; approximate shares; whisker: 95~\% interval, which treats the quasi-random rows as independent). Dashed line: the loading limit; bold: beyond its limit (loading 100~\%, voltage 95–105~\%). Evening is full load without PV; noon is PV output at low load. Some changes come from rounding the number of PV houses.}
\label{fig:7}
\end{figure}

The nine cases are the evening load and two noon PV levels, 5 and 10~\% of houses, on the synthetic feeders of three Sakarya test areas, built from the de-duplicated buildings of Section~5. At noon each PV house carries PV on its whole footprint at 0.14 kWp/m² (E12), and the transformers are sized for the evening load. We varied seven estimated inputs, such as the PV power per roof area and the number of buildings per transformer. The answer is whether any voltage or loading limit is exceeded. We estimate how often it changes from the A and B rows of the feeder's own Sobol sample (base 1,024; 2,048 rows). These rows are quasi-random, so the intervals are indicative.

In Fig.~7, the cases well below the limit hold. The three evening cases, loaded at 58–68~\%, change in at most 1.6~\% of the draws. Suburban noon at 5~\% (25~\%) changes in 5.6~\%, and each of its changes comes with one more PV house. Cases far above the limit hold too: urban noon at 249 and 279~\% changes in at most 3.7~\%. Cases just above the limit do not. Suburban noon at 10~\% (103~\%) changes in 48~\% and rural noon at 5~\% (134~\%) in 56~\%. Rural noon at 10~\% (203~\%) lies between, at 14~\%. In the four cases that change in more than 5~\%, KS puts the PV power per roof area first in three and the number of buildings per transformer first in suburban 5~\%. Sobol points to the same leaders, but its indices did not converge, so we report no index values.

Part of this sensitivity is a modelling artefact. The feeder rounds the number of PV houses per transformer, so an input that sets the cluster size can add or remove one or two PV houses. Close to a limit, that is enough to alter the answer. The building set matters as well. With the first set, which held 22–25 duplicated footprints per area, the shares near the limit differed by up to 18 points, and different cases sat close to it. That an answer near a limit is fragile holds for any threshold. What the nine cases add is how far the fragile zone reached here: up to a loading of 203~\% with the registry values.

\section{District case: Bursa}
Figs.~6 and 7 used representative houses and generic synthetic feeders. What do the three answers look like for one real district? We took 191 houses in Bursa and ran every house (Fig.~8; RQ4). We also asked whether one default system, times 191, gives the same state cost as every house added up.

\begin{figure}[tbp]
\centering
\includegraphics[width=\textwidth]{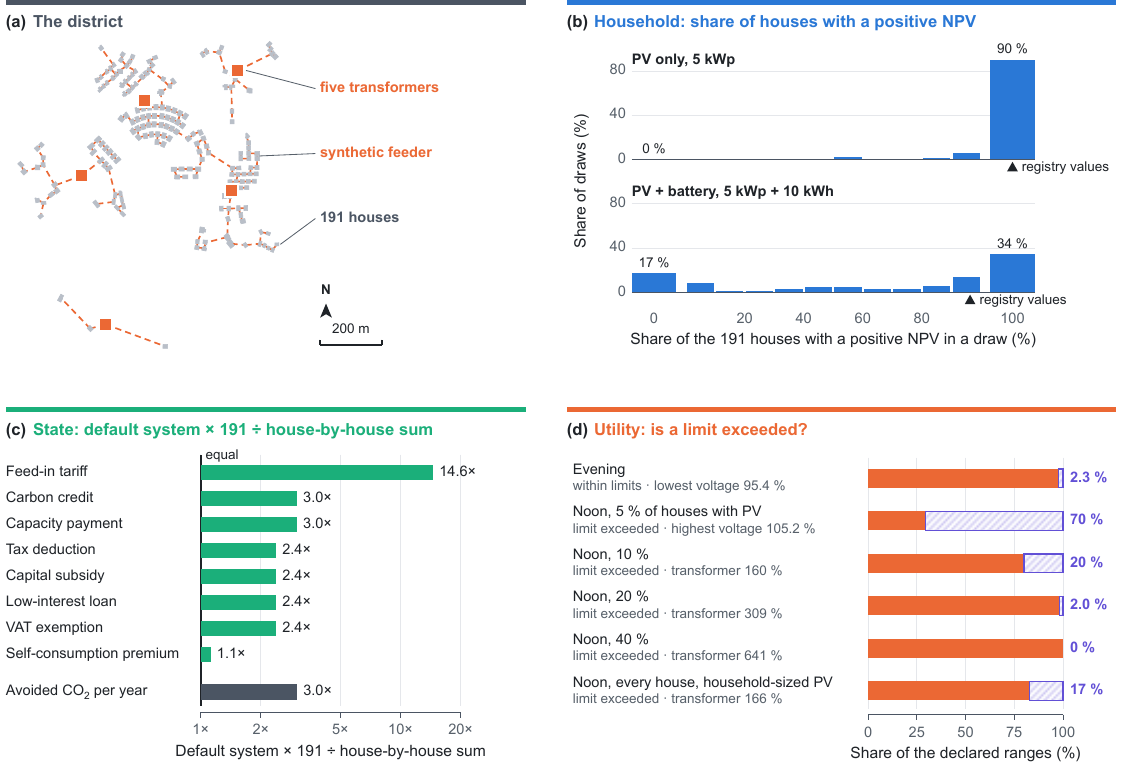}
\caption{The district screening tile of Fig.~1 (``191 real footprints, Bursa'') with its three answers; one demonstration case. (a) Footprints and the synthetic feeder. (b) Draws by the share of houses with a positive NPV, at the prices from 2027 (Section~6.1; post-outcome). (c) The studio's default system × 191 ÷ the house-by-house sum. Green: state cost per incentive; slate: avoided CO\textsubscript{2}. Most ratios equal the size ratio (3.0×; 10 kWp against 3.3 kWp per house on average) or the investment ratio (2.4×); the feed-in tariff follows exports. Net metering is not a candidate. (d) Feeder screening: solid, the same answer as with the registry values; violet hatching, a different answer.}
\label{fig:8}
\end{figure}

A rule fixed before any district was simulated chose the cell. Near the seven weather sites, it took the map cells with the most houses tagged in OpenStreetMap and kept the first in which at least 100 tagged houses matched one footprint each. Four Ankara cells failed, mostly because their houses share merged footprints. A Bursa cell qualified, with 191 houses of 110–362 m² (median 188 m²; Fig.~8a). They use the weather of the Bursa site point and the ranges of Section~6.

A first run lost 25 houses to footprints that the provider returned twice along tile edges. We saw its results, fixed the provider and reran with the frozen rules [post-outcome]. The household conclusions agree; the feeder answers do not.

\textbf{Household.} We gave every house the two systems of Section~6: 5 kWp of PV alone and the same with a 10 kWh battery. Each house uses the energy its footprint area gives (E9), at the corrected prices of Section~6.1 [post-outcome]. With PV alone, every house has a positive NPV with the registry values, and in 90~\% of the draws every house still does (Fig.~8b). With these inputs, an incentive for PV alone would pay only households that invest anyway. Under hourly netting, the median NPV of PV alone falls from 318,000 to 182,000 TL (Table~S3). With the battery, 184 of the 191 houses have a positive NPV; the other seven use less than 3,000 kWh a year and keep the national two-tier tariff. Across the draws the houses often move together: every house invests in 34~\% of the draws and none in 17~\%, while the other 49~\% split. As in Section~6, the discount rate separates these draws most (KS 0.43). The typical dwelling area of E9 comes close (0.42); the district's household draws also vary it (Table~4, last row). All houses share it in a draw, and it sets every house's demand. The supply tariff starts at 3,000 kWh a year, so one draw of the area moves many houses across that limit together.

\textbf{State.} The studio's state view multiplies its default system (10 kWp of PV, a 13 kWh battery, 4,000 kWh a year) by 191; Section~5 used its PV-only form. We compared it with the house-by-house sum, in which each house gets the system the studio sizes for it. The shortcut gives 1.1 to 14.6 times the house-by-house state cost, depending on the incentive, and 3.0 times the avoided CO\textsubscript{2} (Fig.~8c). A decomposition made after the outcome explains most of it [post-outcome]. The houses' own systems total 627 kWp, a third of the default's 1,910.  The size ratio is 3.0 and the investment ratio 2.4. Six of the eight incentive ratios (net metering is not compared), and the CO\textsubscript{2} ratio, equal one of the two. The feed-in tariff (14.6 times) scales with the billed exports, which netting removes for most of these houses, and the self-consumption premium (1.1 times) with the self-consumed energy. In Section~5 the PV-only default was too small for large roofs and understated the cost by 58–87~\%. Here the default is too large for these houses. The error follows the size gap.

\textbf{Utility (feeder).} The 191 houses are 29~\% of the cell's buildings, and the synthetic feeder serves only them, from five transformers. In the evening no limit is exceeded, though the lowest voltage, 0.954 pu, lies close to the 0.95 pu limit; 2.3~\% of the draws differ. At noon, the screening puts 0.14 kWp per m² of footprint on each PV house (E12): a median of 26 kWp per house here, against 5 kWp in the household runs. PV on 5~\% of the houses (ten full roofs) already lifts the highest voltage to 1.052 pu, just over the limit. That answer differs in 70~\% of the draws. With nine PV houses instead of ten, every draw stays under the limit, so the answer turns on a single house.  If every house installs the system the studio sizes for it, one transformer reaches 166~\% and the voltage 1.057 pu; 17~\% of the draws differ (Fig.~8d). In the first run, without the 25 missing houses, the voltages in this case stayed within limits.

The studio can add electric cars, but the frozen runs had none. We added them afterwards, as a demonstration with the registry values where they exist. On today's feeder (Fig.~9a, b), five home-charged cars (3~\% of houses) take the evening voltage below 0.95 pu, and 19 push a transformer over 100~\%. With the voltage only 0.004 pu above the limit, the five-car result depends on which houses get cars. If the studio resizes the transformers, they stay below 100~\%, but from 38 cars the voltage limit still trips. For a household, a car raises the NPV of the same PV by about 128,000 TL (Fig.~9c). Under hourly netting the gain is only 16,000 TL, so monthly netting, which lets midday PV offset the evening charging, causes most of it. With a car, PV with a battery pays for every house under either netting rule.

\begin{figure}[tbp]
\centering
\includegraphics[width=\textwidth]{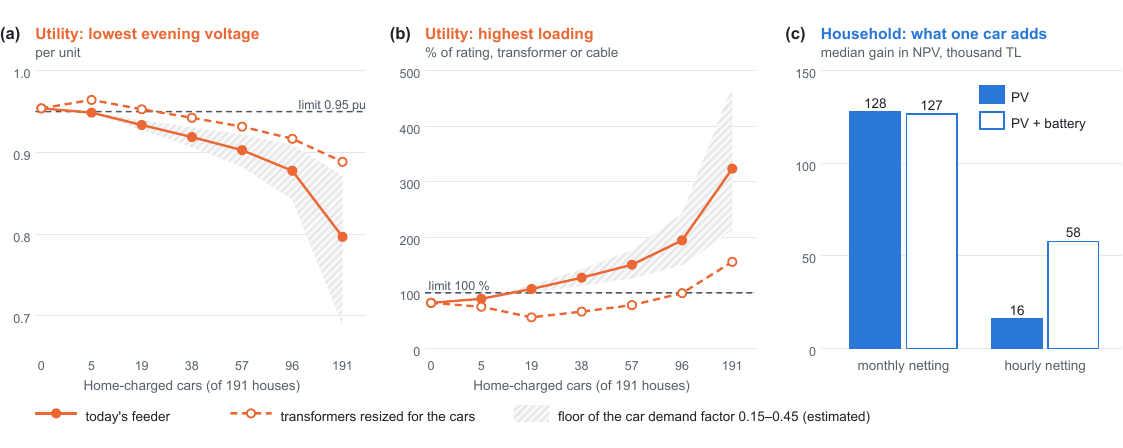}
\caption{Electric cars in the Bursa district, added after the outcome, with the registry values. (a) Lowest evening voltage and (b) highest transformer or cable loading against the number of home-charged cars. Solid: today's feeder; dashed: transformers resized for the cars; grey hatching: today's feeder with the demand-factor floor of the cars from 0.15 to 0.45. Dashed lines: the limits. (c) The median gain in NPV from one car, for the same 5 kWp PV with and without a 10 kWh battery, under monthly and hourly netting, at the prices from 2027.}
\label{fig:9}
\end{figure}

The three answers do not move together. The household answer for PV alone holds in this district. The state cost depends on the system size assumed. The feeder answer turns on single houses and on which buildings are mapped. Cars change the household and feeder answers; the state answer was not run with them. Section~8 collects these limits, with the others the paper has met.

\section{Limitations}
This section lists what the paper does not show, then the choices that shape its numbers.

The platform is not deployed, and no user study was run. In the first runs of their last round, four scripted scenarios show that requests reached the right tools, not that people can use them (Section~3.4). The core was compared with independent tools and one measured national series, not with measured household or feeder data (Section~5). The paper does not evaluate the language model, which is the subject of the companion study (\citealp{kaynak2026}). The MCP path has no guard, so the guard results of that study do not apply to MCP clients.

Several model choices shape the answers. Against the default losses of PVGIS and SAM, the PV yield is 12–16~\% high. In our two cash-flow cases under hourly netting, the degradation rule lowers the NPV by 13–18~\% against SAM. The load shape misses the autumn evening peak, and it was not varied, although the battery answer depends on it. Applying the national tariff's two tiers to the netted monthly imports is our reading of the bill, and one tariff item, the energy fund rate, is unresolved. The supply-tariff price is one twelve-month average of a volatile market price, and the rule does not say whether its 3,000 kWh limit counts consumption before or after netting; we used before. The export price is our reading of the regulation, and we assumed that it holds for supply-tariff customers too. The released studio and chat still use the older prices of Section~6.1, so they give the battery answer ``do not invest''. The model keeps monthly netting after year 10, and Section~7's hourly-netting case bounds that effect (Table~S3). The market price then paid for the excess is not modelled (Section~4). Energy that a battery exports is not paid (\citealp{epdk2026a}); the model's battery does not export, but a meter may not tell the two apart, and then none of the excess is paid. The scenario runs exposed three further choices that were left unchanged: the grid tool's SimBench reference network breaks its voltage limits without PV, the municipal load shape is a synthetic night profile, and the policy tool builds its own simplified cash flow (Table~2, R4).

The sensitivity analysis rests on rule-based ranges, and its shares depend on how wide they are (\citealp{ravalico2010}); the ranges were frozen before the runs, and narrowing the three leading inputs did not remove the changes in the battery answer (Section~6.2). Sourced parameters, such as the tariff and the export price, were held fixed, and where a range is one-sided the registry value lies at its edge. The inputs were drawn independently, so a low discount rate can meet low tariff growth. The loan's market rate (45~\%) is not in the same monetary frame as the discount rate (12~\%).  The state cost is an undiscounted sum (Appendix~A). The studio's feed-in tariff paid on every exported kWh, including those that netting had already credited at the import price. The analyses here and the released code pay it on the billed export only [post-outcome]. This moved the state answer in 0.1–1.4~\% of the draws, always from the feed-in tariff to ``no incentive is enough'', and the Bursa shortcut's feed-in ratio (Section~7) from 8.3 to 14.6. The candidate incentives include the tax deduction, a corporate instrument that a household cannot use; it is among the seven or eight that share the changed draws.

The district is one demonstration case, and its selection favours well-mapped estates. Footprint area stands for dwelling area, so a house with several floors uses more than the model gives it. All houses use the weather of the Bursa site point. At the centre of the district PVGIS gives 5.1~\% more yield, a shift outside the yield range we varied, which only went down.  The electric-car demonstration has no ranges beyond the demand-factor floor of Fig.~9, and its demand-factor constants and charging shape are not in the registry. Its plug-in times follow Swedish data (\citealp{huang2024}), but its hourly shape and its peak per car (2.5–2.9 kW for 5–38 cars, against 0.8–1.0 kW in their averaged profile) are our own. Its demand factor is fitted to a UK design table that we know only from \citet{voronin2025}.

\section{Software and data availability}
\begin{itemize}
\item \textbf{Name:} LAaDT-DSS platform, with the map studio, the network editor, the chat and the MCP server.
\item \textbf{Developer and contact:} Baran Kaynak, Department of Information Systems Engineering, Sakarya University, Sakarya, Türkiye; kaynak@sakarya.edu.tr; ORCID 0000-0002-9004-2639.
\item \textbf{Year first available:} 2026.
\item \textbf{Hardware:} a computer that runs Docker; no GPU is needed for the simulation core.
\item \textbf{Software required:} Docker with Docker Compose. The stack contains Python 3.11, Node.js 20, PostgreSQL and Redis. The chat needs access to a language model; the MCP server runs as a separate container.
\item \textbf{Languages and size:} Python (13,348 lines in 67 files, with 383 test functions) and TypeScript (9,692 lines).
\item \textbf{Availability and licence:} MIT licence. The code, analysis scripts, frozen plans and result files of Sections~5–7 are archived on Zenodo (https://doi.org/10.5281/zenodo.24338714) and mirrored at https://github.com/barankaynak/laadt-dss. The Supplement (Tables~S1–S3) is a separate file submitted with this paper; the result files behind its tables are in the archive.
\item \textbf{Data:} building footprints from the Google–Microsoft open buildings archive (Google Open Buildings, \citealp{sirko2021}, CC BY 4.0; Microsoft Building Footprints, ODbL) and house tags from OpenStreetMap (\citealp{osm2026}; ODbL); typical-year weather from PVGIS (\citealp{huld2012}). The footprint archive is not versioned, so the buildings each analysis used are archived as snapshots, with the PVGIS files and their hashes. The measured national demand (\citealp{epias2026a}) is not redistributed; its hash and download steps are archived.
\item \textbf{Reproduction:} the analysis scripts bypass the platform's result cache. The scenario transcripts of Section~3.4 are archived; the hosted model behind them cannot be rerun. The sensitivity analysis uses SALib (\citealp{herman2017}), the load shapes demandlib 0.1.9 (\citealp{oemof2023}) and the SAM comparison PySAM 7.1.1. The released server's chat runs the corrected orchestrator with the latest guard (v6) by default. The setting \texttt{LLM\_ORCHESTRATOR=frozen} restores the frozen version evaluated in the companion study (Section~3.3).
\end{itemize}

\section{Conclusions}
LAaDT-DSS gives households, the state and distribution utilities their rooftop PV answers from one simulation core, through a map or a chat built for people who are not modellers. The case studies test the system's parts against independent tools and show where its answers are fragile. For a household, PV alone pays in almost every draw, under monthly netting and the supply tariff that applies from 2027. Add a battery, and the answer changes in 31–41~\% of the draws; the state answer, that no incentive is needed, changes with it. Before two prices were corrected after the outcome, the battery answer was ``do not invest'' (Section~6.1). The utility's answer holds far from a limit and is fragile just above it, where single houses and the mapped buildings decide it. Behind these answers, the platform's parts agree with independent tools within stated gaps, and its district shortcuts err mainly with the size of the default system. Four scripted scenarios through the language-model interface found problems that component tests had missed. In the last round, plain requests reached the right tools; whether people without modelling skills can use the studio or the chat is untested.

These change shares come from rule-based ranges and one tariff regime, and the platform has not been compared with measured household or feeder data. Within those bounds, they suggest how such a tool should report. A person who cannot inspect the model sees only the answer, so how often it changes has to stand next to it. The answer for PV alone can then stand as a point answer. The battery and state answers need that change share and their leading inputs next to them. Feeders near a limit need a detailed study. Next steps are a comparison with metered data, the hourly netting that follows the tenth year, and a user study in Turkish.

\section*{CRediT authorship contribution statement}
\textbf{Baran Kaynak:} Conceptualization, Methodology, Software, Validation, Formal analysis, Investigation, Data curation, Writing -- original draft, Writing -- review \& editing, Visualization.

\section*{Funding}
This research did not receive any specific grant from funding agencies in the public, commercial, or not-for-profit sectors.

\section*{Declaration of competing interest}
The author declares no known competing financial interests or personal relationships that could have appeared to influence the work reported in this paper.

\section*{Declaration of generative AI and AI-assisted technologies in the manuscript preparation process}
During the preparation of this work the author used Claude (Anthropic) for software development and to assist in preparing and editing the manuscript text. After using this tool, the author reviewed and edited the content as needed and takes full responsibility for the content of the published article.

\appendix

\section{Model equations as implemented}\label{app:equations}
\begingroup
\setlength{\emergencystretch}{3em}
\setcounter{figure}{0}
\setcounter{table}{0}

\begin{figure}[tbp]
\centering
\includegraphics[width=\textwidth]{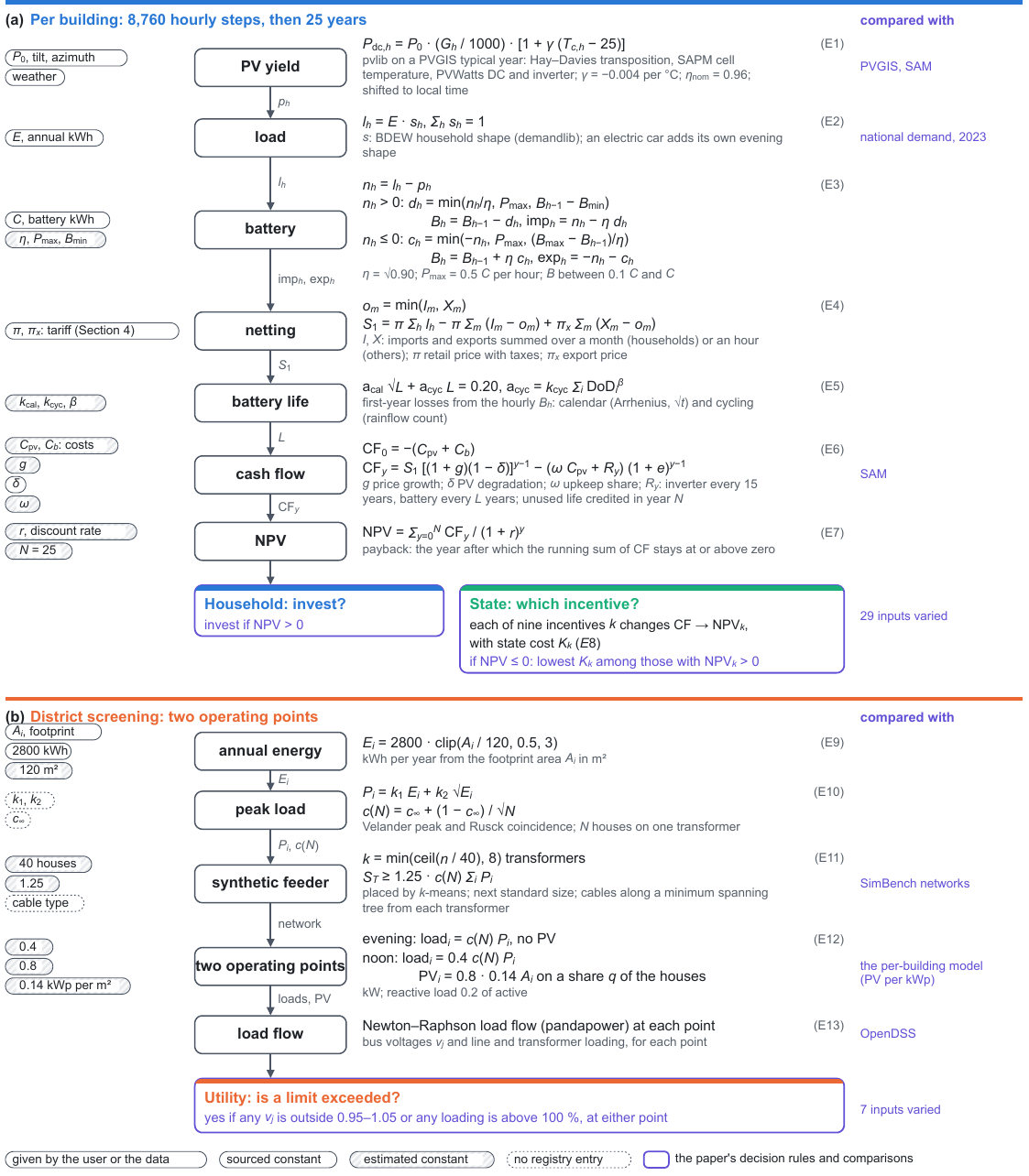}
\caption{The core of Fig.~1, as implemented. (a) Per building: PV yield and load for each hour of a typical
year, the battery rule and its ageing, netting and a 25-year cash flow. (b) District screening: peak loads from
footprint areas, a synthetic feeder and a load flow at two operating points. Hatched: estimated constants; dotted:
no registry entry. Violet: the paper's decision rules, what each step was compared with, and how many inputs were
varied (Section~6). Appendix~A gives all equations.}
\label{fig:A1}
\end{figure}

\subsection{Per building (Fig.~A.1a)}

\noindent\textbf{(E1) PV yield.} \texttt{pv/\allowbreak solar\_\allowbreak model.\allowbreak simulate\_\allowbreak pv\_\allowbreak production}.
\begin{equation}
P_{\mathrm{dc},h} = P_0 \cdot (G_h / 1000) \cdot \bigl[1 + \gamma\,(T_{c,h} - 25)\bigr]
\tag{E1}\label{eq:E1}
\end{equation}
\begin{itemize}
  \item pvlib builds the ModelChain. Its models are inferred from the parameters and printed from the chain:
  \begin{itemize}
    \item transposition: Hay--Davies;
    \item cell temperature: SAPM, open rack, glass--glass;
    \item DC: PVWatts;
    \item inverter: PVWatts, $\eta_{\mathrm{nom}} = 0.96$;
    \item angle-of-incidence loss: physical;
    \item no spectral loss and no extra losses.
  \end{itemize}
  \item $\gamma = -0.004$ per \textdegree C.
  \item The weather is a PVGIS typical year, cached per site with its SHA-256.
  \item The 8,760 values are rolled by +3 hours to local time. The roll wraps three hours across the year boundary.
\end{itemize}

\noindent\textbf{(E2) Load.} \texttt{loads/\allowbreak standard\_\allowbreak profiles.\allowbreak bdew\_\allowbreak profile}.
\begin{equation}
l_h = E \cdot s_h, \quad \text{with} \quad \sum_h s_h = 1.
\tag{E2}\label{eq:E2}
\end{equation}
\begin{itemize}
  \item $s$ is the BDEW household shape from demandlib: year 2023, no holidays, 15-minute values summed to hours, then
  divided by their sum.
  \item An electric car adds its own evening profile (\texttt{loads/\allowbreak ev\_\allowbreak load.\allowbreak ev\_\allowbreak charging\_\allowbreak profile}), but only when its annual energy
  is above zero.
  \item The municipal shape is a hand-written night profile, not BDEW. Fig.~A.1 shows the household case only.
\end{itemize}

\noindent\textbf{(E3) Battery.} \texttt{bess/\allowbreak battery\_\allowbreak model.\allowbreak dispatch\_\allowbreak self\_\allowbreak consumption}.
\begin{equation}
n_h = l_h - p_h.
\tag{E3}\label{eq:E3}
\end{equation}
\begin{itemize}
  \item If $n_h > 0$:
  \begin{itemize}
    \item $d_h = \min\bigl(n_h/\eta,\ P_{\max},\ B_{h-1} - B_{\min}\bigr)$;
    \item $B_h = B_{h-1} - d_h$;
    \item $\mathrm{imp}_h = n_h - \eta\, d_h$.
  \end{itemize}
  \item If $n_h \le 0$:
  \begin{itemize}
    \item $c_h = \min\bigl(-n_h,\ P_{\max},\ (B_{\max} - B_{h-1})/\eta\bigr)$;
    \item $B_h = B_{h-1} + \eta\, c_h$;
    \item $\mathrm{exp}_h = -n_h - c_h$.
  \end{itemize}
  \item $\eta = \sqrt{0.90}$ per direction, and $P_{\max} = 0.5\,C$ per hour. $B$ stays between $0.1\,C$ and $C$, and the battery starts at
  $B_{\min}$.
  \item Discharge is counted on the battery side: $n/\eta$ is requested and $\eta\, d$ is delivered.
  \item The dispatch does not lose capacity over the years. Ageing only sets the replacement year in (E6).
\end{itemize}

\noindent\textbf{(E4) Netting.} \texttt{economic/\allowbreak economic\_\allowbreak engine.\allowbreak apply\_\allowbreak netting} and \texttt{simulate\_\allowbreak annual\_\allowbreak cashflow}, with the tariff from
\texttt{tariffs/\allowbreak turkey\_\allowbreak 2026.\allowbreak get\_\allowbreak simple\_\allowbreak tariff}.
\begin{align}
o_m &= \min(I_m, X_m) \notag\\
B_m(x) &= \pi_1 \min(x, 8 d_m) + \pi_2 \max(0, x - 8 d_m) \notag\\
S_1 &= \sum_m \big[ B_m(L_m) - B_m(I_m - o_m) \big] + \pi_x \sum_m (X_m - o_m)
\tag{E4}\label{eq:E4}
\end{align}
\begin{itemize}
  \item $I$ and $X$ are imports and exports summed over a netting period. The period is a month for households and an hour for
  the others (\texttt{netting\_\allowbreak interval\_\allowbreak for}).
  \item $L_m$ is the month's load and $d_m$ its number of days. On the national tariff $\pi_1 = 3.238$ and
  $\pi_2 = 4.857$ TL/kWh with taxes: the first 8 kWh a day of each month at the first tier, the rest at the second
  (applied to the netted imports after the outcome, \texttt{scripts/\allowbreak tier}; the frozen runs used
  $\pi_1 = \pi_2 = 4.857$). From 2027, households using 3,000 kWh a year or more pay the last-resort supply
  tariff, $\pi_1 = \pi_2 = 6.096$ TL/kWh (Section~4). Under hourly netting (Section~7's check), $I$ and $X$
  are netted per hour and then summed per month. $\pi_x = 1.8958$ TL/kWh, the high-tier active-energy price; the studio and the frozen runs used 0.4941.
  \item Netting is applied after the hourly dispatch.
\end{itemize}

\noindent\textbf{(E5) Battery life.} \texttt{bess/\allowbreak degradation.\allowbreak calculate\_\allowbreak annual\_\allowbreak degradation} and \texttt{economic/\allowbreak lifecycle.\allowbreak battery\_\allowbreak life\_\allowbreak years}.
\begin{equation}
a_{\mathrm{cal}} \sqrt{L} + a_{\mathrm{cyc}} L = 0.20, \quad \text{with} \quad a_{\mathrm{cyc}} = k_{\mathrm{cyc}} \sum_i \mathrm{DoD}_i^{\beta}.
\tag{E5}\label{eq:E5}
\end{equation}
\begin{itemize}
  \item The first-year losses come from the hourly $B_h$. The calendar term follows Arrhenius and $\sqrt{t}$; the cycling term uses a
  rainflow count.
  \item The temperature in this path is the default 25\,\textdegree C, so the Arrhenius factor is 1.
  \item The calendar term carries an SOC factor $\operatorname{clip}\bigl(1 + 0.3\,(\text{mean SOC} - 0.5),\ 0.5,\ 2\bigr)$.
  \item Cycles shallower than 1\,\% are ignored. Leftover half-cycles enter as $(0.5 \cdot \mathrm{DoD})^{\beta}$.
  \item The two first-year losses are rounded to four decimals before $L$ is solved.
\end{itemize}

\noindent\textbf{(E6) Cash flow.} \texttt{economic/\allowbreak lifecycle.\allowbreak build\_\allowbreak cash\_\allowbreak flows}.
\begin{align}
\mathrm{CF}_0 &= -(C_{\mathrm{pv}} + C_{\mathrm{b}}) \notag\\
\mathrm{CF}_y &= S_1 \bigl[(1 + g)(1 - \delta)\bigr]^{y-1} - (\omega\, C_{\mathrm{pv}} + R_y)(1 + e)^{y-1}
\tag{E6}\label{eq:E6}
\end{align}
\begin{itemize}
  \item Degradation multiplies the whole savings, not the energy. It is the main cause of the gap to SAM (Section~5).
  \item $e = g$ unless it is set.
  \item $R_y$ is the replacement cost: the inverter every 15 years and the battery every $L$ years, in the years $\lceil kL \rceil < N$.
  \item The unused life of a component is credited in year $N$.
  \item $\omega$ is \texttt{PV\_\allowbreak OM\_\allowbreak FRACTION\_\allowbreak OF\_\allowbreak CAPEX\_\allowbreak PER\_\allowbreak YEAR}.
\end{itemize}

\noindent\textbf{(E7) NPV and payback.} \texttt{economic/\allowbreak economic\_\allowbreak engine.\allowbreak calculate\_\allowbreak npv} and \texttt{economic/\allowbreak lifecycle.\allowbreak payback\_\allowbreak years}.
\begin{equation}
\mathrm{NPV} = \sum_{y=0}^{N} \frac{\mathrm{CF}_y}{(1 + r)^y}, \quad \text{with} \quad N = 25.
\tag{E7}\label{eq:E7}
\end{equation}
\begin{itemize}
  \item Payback is the year after which the running sum of CF stays at or above zero. This is the last crossing, not the
  first.
  \item The older \texttt{calculate\_\allowbreak payback\_\allowbreak period} takes the first crossing; the policy engine uses it.
  \item $\mathrm{LCOE} = C_0 \big/ \bigl(P \sum_{y=1}^{N} (1+r)^{-y}\bigr)$, with $P$ the first-year production (as in E8a). It includes no upkeep and no degradation, and it is not in Fig.~A.1.
\end{itemize}

\noindent\textbf{(E8) The paper's decision rules} (not the studio's; \texttt{scripts/\allowbreak rq3/\allowbreak model.py}).
\begin{itemize}
  \item \textbf{Household:} invest if $\mathrm{NPV} > 0$. The studio words its own verdict on the same threshold
  (\texttt{advisor.\allowbreak \_plain\_\allowbreak language}).
  \item \textbf{State:} if $\mathrm{NPV} \le 0$, choose the lowest state cost $K_k$ among the incentives with $\mathrm{NPV}_k > 0$ (\texttt{choose\_\allowbreak incentive}).
  \begin{itemize}
    \item Net metering is not a candidate.
    \item Incentives with zero cost are excluded.
    \item The studio itself reports ``best by NPV'' and ``best by added NPV per TL of state cost'' (\texttt{advisor.\allowbreak policy}).
  \end{itemize}
\end{itemize}

\noindent\textbf{(E8a) The nine incentive mechanisms.} (E8) needs an $\mathrm{NPV}_k$ and a state cost $K_k$ for each incentive. Both come
from the nine mechanisms below: \texttt{policy/\allowbreak policy\_\allowbreak engine.\allowbreak apply\_\allowbreak incentive} changes the cash flow of (E6), and
\texttt{api/\allowbreak routes/\allowbreak advisor.\allowbreak \_policy\_\allowbreak state\_\allowbreak cost} gives $K_k$. Table~A.1 lists them. The levels are the studio's \texttt{DEFAULT\_\allowbreak POLICIES}; each level is a
registry entry marked estimated (\texttt{policy.\allowbreak studio.*}).
\begin{itemize}
  \item $C_0 = -\mathrm{CF}_0$ is the investment. The code assumes it includes 20\,\% VAT.
  \item $X^{\mathrm{g}}$ is the first-year export before netting (\texttt{total\_\allowbreak grid\_\allowbreak export\_\allowbreak kwh}). $X^{\mathrm{b}} = \sum_m (X_m - o_m)$ is the billed export after netting (E4).
  The feed-in tariff pays only for $X^{\mathrm{b}}$, since netting already credits the netted export at the import price.
  The studio paid it on $X^{\mathrm{g}}$ until a correction [post-outcome] (Section~8). $P$ is the
  first-year production, $Q = \max(0, P - X^{\mathrm{g}})$ the self-consumed energy, $\pi_x$ the export price, $P_{\mathrm{kWp}}$ the rated power
  and $N = 25$ years.
  \item $X^{\mathrm{g}}$, $P$ and $Q$ keep their first-year values in every year. ``$\mathrm{CF}_y \mathrel{+}=$'' adds the same amount in each year $y = 1\dots N$, so
  the incentive terms do not degrade, while the savings in (E6) do.
  \item $A(D, r, n) = D\, r \big/ \bigl(1 - (1 + r)^{-n}\bigr)$ is the annual repayment of a loan $D$ at rate $r$ over $n$ years.
\end{itemize}

\begin{table}[tbp]
\centering
\footnotesize
\caption{The nine incentive mechanisms of (E8a): change to the cash flow and state cost per system.}\label{tab:A1}
\begin{tabularx}{\linewidth}{@{}>{\raggedright\arraybackslash}X>{\raggedright\arraybackslash}X>{\raggedright\arraybackslash}X@{}}
\toprule
Mechanism (level) & Change to the cash flow & State cost $K_k$ per system \\
\midrule
Capital subsidy (30\,\%) & $\mathrm{CF}_0 \gets 0.70\,\mathrm{CF}_0$ & $0.30\,C_0$ \\
VAT exemption (20\,\%) & $\mathrm{CF}_0 \mathrel{+}= 0.20\,C_0 / 1.20$ & $0.20\,C_0 / 1.20$ \\
Tax deduction (100\,\%, corporate rate 25\,\%) & $\mathrm{CF}_0 \mathrel{+}= 1.00 \cdot 0.25\,C_0$ & $0.25\,C_0$ \\
Net metering (1:1) & none at the current ratio of 1 & $X^{\mathrm{g}} \pi \cdot \max(0, 1 - 1) \cdot N = 0$ \\
Feed-in tariff (7 TL/kWh) & $\mathrm{CF}_y \mathrel{+}= X^{\mathrm{b}} (7 - \pi_x)$ & $X^{\mathrm{b}} \cdot \max(0, 7 - \pi_x) \cdot N$ \\
Capacity payment (200 TL/kW/yr) & $\mathrm{CF}_y \mathrel{+}= 200\,P_{\mathrm{kWp}}$ & $200\,P_{\mathrm{kWp}}\,N$ \\
Low-interest loan (80\,\% of $C_0$, 15\,\%, 10 years; market rate 45\,\%) & $\mathrm{CF}_0 \mathrel{+}= D$; $\mathrm{CF}_y \mathrel{-}= A(D, 0.15, 10)$ for $y = 1\dots 10$, with $D = 0.80\,C_0$ & $D\,(0.45 - 0.15) \cdot 10$ \\
Self-consumption premium (0.5 TL/kWh) & $\mathrm{CF}_y \mathrel{+}= 0.5\,Q$ & $0.5\,Q\,N$ \\
Carbon credit (500 TL/t) & $\mathrm{CF}_y \mathrel{+}= 500 \cdot 0.45\,P / 1000$ & $500 \cdot 0.45\,P\,N / 1000$ \\
\bottomrule
\end{tabularx}
\end{table}

The market rate enters only $K_k$. Net metering costs the state nothing at a ratio of 1, so (E8) drops it. $K_k$ is a
plain sum over the years, with no discounting; for the loan it is the rate spread times the principal times the
term. This charges the spread on the full principal for all ten years, as if the loan were interest-only; for a loan
that is paid down it is an upper bound. At the registry rates the loan lowers the household's NPV: the repayments,
discounted at 12\,\%, exceed the principal, so $\Delta\mathrm{NPV} \approx -0.126\,D$. It wins the state answer in
one of the 2,048 draws of Section~6, at Sakarya. Discounting $K_k$ at 3\,\% or 12\,\% changes the state answer of (E8) in at most 2.6\,\% of the draws of the frozen runs of Section~6, and
not with the registry values. The tax deduction assumes a corporate tax payer; a household cannot use it (Section~8).

These nine terms act on one building. The equations below scale up to a district.

\subsection{District screening (Fig.~A.1b)}

\noindent\textbf{(E9) Annual energy.} \texttt{loads/\allowbreak turkey\_\allowbreak parameters.\allowbreak annual\_\allowbreak kwh\_\allowbreak from\_\allowbreak area}.
\begin{equation}
E_i = 2800 \cdot \operatorname{clip}(A_i / 120,\ 0.5,\ 3)
\tag{E9}\label{eq:E9}
\end{equation}
\begin{itemize}
  \item The clamp at 3.0 is why the district case keeps to low-rise houses.
\end{itemize}

\noindent\textbf{(E10) Peak and coincidence.} \texttt{loads/\allowbreak demand\_\allowbreak model.\allowbreak velander\_\allowbreak peak\_\allowbreak kw} and \texttt{coincidence\_\allowbreak factor}.
\begin{align}
P_i &= k_1 E_i + k_2 \sqrt{E_i} \notag\\
c(N) &= c_\infty + (1 - c_\infty)/\sqrt{N}
\tag{E10}\label{eq:E10}
\end{align}
\begin{itemize}
  \item $k_1 = 0.33 \times 10^{-3}$ kW per kWh, $k_2 = 0.05$ kW per $\sqrt{\mathrm{kWh}}$ and $c_\infty = 0.10$, with $E_i$ in kWh a year and $P_i$ in kW.
  \item \textbf{None of these three has an entry in the parameter registry.} \texttt{demand\_\allowbreak model.py} is not among the modules the
  registry test covers.
\end{itemize}

\noindent\textbf{(E11) Synthetic feeder.} \texttt{advisor/\allowbreak region\_\allowbreak network.\allowbreak \_auto\_\allowbreak place\_\allowbreak transformers}, \texttt{\_build\_\allowbreak mst} and
\texttt{\_pick\_\allowbreak trafo\_\allowbreak rating\_\allowbreak kva}.
\begin{itemize}
  \item $k = \min(\lceil n/40 \rceil, 8)$ transformers.
  \item $S_T \ge 1.25 \cdot c(N) \sum_i P_i$, rounded up to the next standard size. $S_T$ also covers the EV term when the EV share is
  above zero, and is at least 100 kVA.
  \item k-means is the platform's own code: farthest-point start, at most six rounds.
  \item Cables follow a minimum spanning tree from each transformer. The type is NAYY 4x150 SE, the shortest segment is
  20 m, and lengths are great-circle.
  \item $n$ is the number of all houses; $N$ is the number of houses on one transformer.
  \item The cable type has no registry entry.
\end{itemize}

\noindent\textbf{(E12) Two operating points.} \texttt{advisor/\allowbreak region\_\allowbreak network.\allowbreak \_build\_\allowbreak and\_\allowbreak run}.
\begin{itemize}
  \item Evening: $\mathrm{load}_i = c(N)\, P_i$, no PV.
  \item Noon: $\mathrm{load}_i = 0.4\, c(N)\, P_i$, and $\mathrm{PV}_i = 0.8 \cdot 0.14\, A_i$ on a share $q$ of the houses.
  \item Loads are in kW; reactive load is 0.2 of active load.
  \item The PV houses are the first $\operatorname{round}(qN)$ houses in each transformer's list, not a random draw.
  \item With storage, the code scales the loads and the PV by fixed factors (0.3 and 0.5). The frozen runs of Sections~6--7 use no storage and no electric cars; Section~7 adds a car demonstration, also without storage.
  \item Electric cars (\texttt{loads/\allowbreak ev\_\allowbreak load}): in the evening, each house with a car adds $7.4\,f(n)$ kW, with
  $f(n) = c_{\mathrm{ev}} + (0.50 - c_{\mathrm{ev}})/\sqrt{n} = 0.30 + 0.20/\sqrt{n}$ for $n$ cars on one transformer. Section~7
  varies $c_{\mathrm{ev}}$ from 0.15 to 0.45. In the household cash flow a car adds 3,000 kWh a year with its evening shape (E2).
  The annual energy (sourced) and the charger power (estimated) are registry entries; $c_{\mathrm{ev}}$, the 0.50 and the
  charging shape are not.
  \item A fixed yield of 1,400 kWh per kWp (\texttt{advisor/\allowbreak sizing.\allowbreak SPECIFIC\_\allowbreak YIELD\_\allowbreak KWH\_\allowbreak PER\_\allowbreak KWP}) sizes a
  house's kWp from its annual energy. It sets the studio-sized systems of Section~7, also on the feeder
  (\texttt{scripts/\allowbreak rq4/\allowbreak feeder.py}). The screening comparison of Section~5 uses it for the annual production of
  area-sized systems (\texttt{scripts/\allowbreak verify/\allowbreak screening.py}).
\end{itemize}

\noindent\textbf{(E13) Load flow.} \texttt{pp.runpp} in \texttt{\_build\_\allowbreak and\_\allowbreak run}; \texttt{simulate\_\allowbreak region} merges the results.
\begin{itemize}
  \item pandapower solves the load flow with Newton--Raphson, its default.
  \item Every transformer has its own 10.5 kV slack bus at 1.0 pu, so the feeder is several independent sub-networks.
  \item \textbf{The paper's rule} (\texttt{feeder\_\allowbreak decisions} and \texttt{cell\_\allowbreak outputs}): at each operating point on its own, the limit is
  exceeded if any bus voltage is outside 0.95--1.05 pu, a planning band tighter than the $\pm$10\,\% of EN 50160
  (\citealp{cenelec2022}), or any line or transformer loading is above 100\,\%. A load flow
  that does not converge counts as exceeded.
  \item \textbf{The studio merges the two points} before it reports a result:
  \begin{itemize}
    \item the lowest voltage comes from the evening;
    \item the highest voltage and the loading come from both points;
    \item only building buses are used;
    \item its overload flag reads transformers only.
  \end{itemize}
\end{itemize}
\endgroup

\bibliographystyle{elsarticle-harv}
\bibliography{references}

\begin{thebibliography}{76}
\expandafter\ifx\csname natexlab\endcsname\relax\def\natexlab#1{#1}\fi
\providecommand{\url}[1]{\texttt{#1}}
\providecommand{\href}[2]{#2}
\providecommand{\path}[1]{#1}
\providecommand{\DOIprefix}{doi:}
\providecommand{\ArXivprefix}{arXiv:}
\providecommand{\URLprefix}{URL: }
\providecommand{\Pubmedprefix}{pmid:}
\providecommand{\doi}[1]{\href{http://dx.doi.org/#1}{\path{#1}}}
\providecommand{\Pubmed}[1]{\href{pmid:#1}{\path{#1}}}
\providecommand{\bibinfo}[2]{#2}
\ifx\xfnm\relax \def\xfnm[#1]{\unskip,\space#1}\fi
\bibitem[{Anderson et~al.(2023)Anderson, Hansen, Holmgren, Jensen, Mikofski and
  Driesse}]{anderson2023}
\bibinfo{author}{Anderson, K.S.}, \bibinfo{author}{Hansen, C.W.},
  \bibinfo{author}{Holmgren, W.F.}, \bibinfo{author}{Jensen, A.R.},
  \bibinfo{author}{Mikofski, M.A.}, \bibinfo{author}{Driesse, A.},
  \bibinfo{year}{2023}.
\newblock \bibinfo{title}{{pvlib python: 2023 project update}}.
\newblock \bibinfo{journal}{Journal of Open Source Software}
  \bibinfo{volume}{8}, \bibinfo{pages}{5994}.
\newblock \DOIprefix\doi{10.21105/joss.05994}.
\bibitem[{Ardila-Cruz et~al.(2027)Ardila-Cruz, Pineda-Rangel and
  Mu{\~n}oz~Maldonado}]{ardilacruz2027}
\bibinfo{author}{Ardila-Cruz, S.}, \bibinfo{author}{Pineda-Rangel, S.},
  \bibinfo{author}{Mu{\~n}oz~Maldonado, Y.A.}, \bibinfo{year}{2027}.
\newblock \bibinfo{title}{{Financial profitability of commercial solar on-grid
  PV in Colombia: A case study on the impact of fiscal incentives}}.
\newblock \bibinfo{journal}{Renewable Energy Focus} \bibinfo{volume}{60},
  \bibinfo{pages}{100939}.
\newblock \DOIprefix\doi{10.1016/j.ref.2026.100939}.
\bibitem[{Becchi et~al.(2025)Becchi, Bindi, Intravaia, Lozito, Piccirilli,
  Marconi and Cacciavillani}]{becchi2025}
\bibinfo{author}{Becchi, L.}, \bibinfo{author}{Bindi, M.},
  \bibinfo{author}{Intravaia, M.}, \bibinfo{author}{Lozito, G.M.},
  \bibinfo{author}{Piccirilli, M.C.}, \bibinfo{author}{Marconi, R.},
  \bibinfo{author}{Cacciavillani, V.}, \bibinfo{year}{2025}.
\newblock \bibinfo{title}{{GPT-Powered Chatbot for Dissemination and Simulation
  of Renewable Energy Communities}}, in: \bibinfo{booktitle}{2025 IEEE
  International Conference on Environment and Electrical Engineering and 2025
  IEEE Industrial and Commercial Power Systems Europe (EEEIC / I\&CPS Europe)},
  pp. \bibinfo{pages}{1--6}.
\newblock \DOIprefix\doi{10.1109/EEEIC/ICPSEurope64998.2025.11169160}.
\bibitem[{Blair et~al.(2018)Blair, DiOrio, Freeman, Gilman, Janzou, Neises and
  Wagner}]{blair2018}
\bibinfo{author}{Blair, N.}, \bibinfo{author}{DiOrio, N.},
  \bibinfo{author}{Freeman, J.}, \bibinfo{author}{Gilman, P.},
  \bibinfo{author}{Janzou, S.}, \bibinfo{author}{Neises, T.},
  \bibinfo{author}{Wagner, M.}, \bibinfo{year}{2018}.
\newblock \bibinfo{title}{{System Advisor Model (SAM) General Description
  (Version 2017.9.5)}}.
\newblock \bibinfo{type}{Technical Report}
  \bibinfo{number}{NREL/TP-6A20-70414}. National Renewable Energy Laboratory.
  \bibinfo{address}{Golden, CO}.
\newblock \DOIprefix\doi{10.2172/1440404}.
\bibitem[{Boulaire et~al.(2015)Boulaire, Utting and Drogemuller}]{boulaire2015}
\bibinfo{author}{Boulaire, F.}, \bibinfo{author}{Utting, M.},
  \bibinfo{author}{Drogemuller, R.}, \bibinfo{year}{2015}.
\newblock \bibinfo{title}{{Impact of technology uptake on an Australian
  electricity distribution network}}.
\newblock \bibinfo{journal}{Environmental Modelling \& Software}
  \bibinfo{volume}{69}, \bibinfo{pages}{196--213}.
\newblock \DOIprefix\doi{10.1016/j.envsoft.2015.03.019}.
\bibitem[{Branger et~al.(2015)Branger, Giraudet, Guivarch and
  Quirion}]{branger2015}
\bibinfo{author}{Branger, F.}, \bibinfo{author}{Giraudet, L.G.},
  \bibinfo{author}{Guivarch, C.}, \bibinfo{author}{Quirion, P.},
  \bibinfo{year}{2015}.
\newblock \bibinfo{title}{{Global sensitivity analysis of an energy--economy
  model of the residential building sector}}.
\newblock \bibinfo{journal}{Environmental Modelling \& Software}
  \bibinfo{volume}{70}, \bibinfo{pages}{45--54}.
\newblock \DOIprefix\doi{10.1016/j.envsoft.2015.03.021}.
\bibitem[{Brinkerink et~al.(2022)Brinkerink, Zakeri, Huppmann, Glynn,
  {\'O}~Gallach{\'o}ir and Deane}]{brinkerink2022}
\bibinfo{author}{Brinkerink, M.}, \bibinfo{author}{Zakeri, B.},
  \bibinfo{author}{Huppmann, D.}, \bibinfo{author}{Glynn, J.},
  \bibinfo{author}{{\'O}~Gallach{\'o}ir, B.}, \bibinfo{author}{Deane, P.},
  \bibinfo{year}{2022}.
\newblock \bibinfo{title}{{Assessing global climate change mitigation scenarios
  from a power system perspective using a novel multi-model framework}}.
\newblock \bibinfo{journal}{Environmental Modelling \& Software}
  \bibinfo{volume}{150}, \bibinfo{pages}{105336}.
\newblock \DOIprefix\doi{10.1016/j.envsoft.2022.105336}.
\bibitem[{{\c C}e{\c c}en et~al.(2022){\c C}e{\c c}en, Yavuz, T{\i}rm{\i}k{\c
  c}{\i}, Sar{\i}kaya and Yan{\i}ko{\u g}lu}]{cecen2022}
\bibinfo{author}{{\c C}e{\c c}en, M.}, \bibinfo{author}{Yavuz, C.},
  \bibinfo{author}{T{\i}rm{\i}k{\c c}{\i}, C.A.}, \bibinfo{author}{Sar{\i}kaya,
  S.}, \bibinfo{author}{Yan{\i}ko{\u g}lu, E.}, \bibinfo{year}{2022}.
\newblock \bibinfo{title}{{Analysis and evaluation of distributed photovoltaic
  generation in electrical energy production and related regulations of
  Turkey}}.
\newblock \bibinfo{journal}{Clean Technologies and Environmental Policy}
  \bibinfo{volume}{24}, \bibinfo{pages}{1321--1336}.
\newblock \DOIprefix\doi{10.1007/s10098-021-02247-0}.
\bibitem[{{CENELEC}(2022)}]{cenelec2022}
\bibinfo{author}{{CENELEC}}, \bibinfo{year}{2022}.
\newblock \bibinfo{title}{{{EN} 50160:2022. Voltage characteristics of
  electricity supplied by public electricity networks}}.
\newblock \bibinfo{howpublished}{European Committee for Electrotechnical
  Standardization, Brussels}.
\bibitem[{Chappin et~al.(2017)Chappin, de~Vries, Richstein, Bhagwat, Iychettira
  and Khan}]{chappin2017}
\bibinfo{author}{Chappin, E.J.}, \bibinfo{author}{de~Vries, L.J.},
  \bibinfo{author}{Richstein, J.C.}, \bibinfo{author}{Bhagwat, P.},
  \bibinfo{author}{Iychettira, K.}, \bibinfo{author}{Khan, S.},
  \bibinfo{year}{2017}.
\newblock \bibinfo{title}{{Simulating climate and energy policy with
  agent-based modelling: The Energy Modelling Laboratory (EMLab)}}.
\newblock \bibinfo{journal}{Environmental Modelling \& Software}
  \bibinfo{volume}{96}, \bibinfo{pages}{421--431}.
\newblock \DOIprefix\doi{10.1016/j.envsoft.2017.07.009}.
\bibitem[{Charan et~al.(2021)Charan, Mackey, Irani, Polly, Ray, Fleming,
  El~Kontar, Moore, Elgindy, Cutler, Roudsari and Goldwasser}]{charan2021}
\bibinfo{author}{Charan, T.}, \bibinfo{author}{Mackey, C.},
  \bibinfo{author}{Irani, A.}, \bibinfo{author}{Polly, B.},
  \bibinfo{author}{Ray, S.}, \bibinfo{author}{Fleming, K.},
  \bibinfo{author}{El~Kontar, R.}, \bibinfo{author}{Moore, N.},
  \bibinfo{author}{Elgindy, T.}, \bibinfo{author}{Cutler, D.},
  \bibinfo{author}{Roudsari, M.S.}, \bibinfo{author}{Goldwasser, D.},
  \bibinfo{year}{2021}.
\newblock \bibinfo{title}{{Integration of Open-Source {URBANopt} and
  {Dragonfly} Energy Modeling Capabilities into Practitioner Workflows for
  District-Scale Planning and Design}}.
\newblock \bibinfo{journal}{Energies} \bibinfo{volume}{14},
  \bibinfo{pages}{5931}.
\newblock \DOIprefix\doi{10.3390/en14185931}.
\bibitem[{Cutler et~al.(2017)Cutler, Olis, Elgqvist, Li, Laws, DiOrio, Walker
  and Anderson}]{cutler2017}
\bibinfo{author}{Cutler, D.}, \bibinfo{author}{Olis, D.},
  \bibinfo{author}{Elgqvist, E.}, \bibinfo{author}{Li, X.},
  \bibinfo{author}{Laws, N.}, \bibinfo{author}{DiOrio, N.},
  \bibinfo{author}{Walker, A.}, \bibinfo{author}{Anderson, K.},
  \bibinfo{year}{2017}.
\newblock \bibinfo{title}{{REopt: A Platform for Energy System Integration and
  Optimization}}.
\newblock \bibinfo{type}{Technical Report}
  \bibinfo{number}{NREL/TP-7A40-70022}. National Renewable Energy Laboratory.
  \bibinfo{address}{Golden, CO}.
\newblock \DOIprefix\doi{10.2172/1395453}.
\bibitem[{De~Groote and Verboven(2019)}]{degroote2019}
\bibinfo{author}{De~Groote, O.}, \bibinfo{author}{Verboven, F.},
  \bibinfo{year}{2019}.
\newblock \bibinfo{title}{{Subsidies and Time Discounting in New Technology
  Adoption: Evidence from Solar Photovoltaic Systems}}.
\newblock \bibinfo{journal}{American Economic Review} \bibinfo{volume}{109},
  \bibinfo{pages}{2137--2172}.
\newblock \DOIprefix\doi{10.1257/aer.20161343}.
\bibitem[{Dengler et~al.(2023)Dengler, Bazan, German, Lalbakhsh and
  Liebmann}]{dengler2023}
\bibinfo{author}{Dengler, G.}, \bibinfo{author}{Bazan, P.},
  \bibinfo{author}{German, R.}, \bibinfo{author}{Lalbakhsh, P.},
  \bibinfo{author}{Liebmann, A.}, \bibinfo{year}{2023}.
\newblock \bibinfo{title}{{A Conversational Human-Computer Interface for Smart
  Energy System Simulation Environments}}, in: \bibinfo{booktitle}{2023 Winter
  Simulation Conference (WSC)}, pp. \bibinfo{pages}{2978--2989}.
\newblock \DOIprefix\doi{10.1109/WSC60868.2023.10408707}.
\bibitem[{Dickert and Schegner(2010)}]{dickert2010}
\bibinfo{author}{Dickert, J.}, \bibinfo{author}{Schegner, P.},
  \bibinfo{year}{2010}.
\newblock \bibinfo{title}{{Residential load models for network planning
  purposes}}, in: \bibinfo{booktitle}{Proceedings of the 2010 International
  Symposium on Modern Electric Power Systems (MEPS), Wroc{\l}aw, Poland}, pp.
  \bibinfo{pages}{1--6}.
\bibitem[{Dobos(2014)}]{dobos2014}
\bibinfo{author}{Dobos, A.P.}, \bibinfo{year}{2014}.
\newblock \bibinfo{title}{{PVWatts Version 5 Manual}}.
\newblock \bibinfo{type}{Technical Report}
  \bibinfo{number}{NREL/TP-6A20-62641}. National Renewable Energy Laboratory.
  \bibinfo{address}{Golden, CO}.
\newblock \DOIprefix\doi{10.2172/1158421}.
\bibitem[{Dufo-L{\'o}pez and Bernal-Agust{\'i}n(2015)}]{dufolopez2015}
\bibinfo{author}{Dufo-L{\'o}pez, R.}, \bibinfo{author}{Bernal-Agust{\'i}n,
  J.L.}, \bibinfo{year}{2015}.
\newblock \bibinfo{title}{{A comparative assessment of net metering and net
  billing policies. Study cases for Spain}}.
\newblock \bibinfo{journal}{Energy} \bibinfo{volume}{84},
  \bibinfo{pages}{684--694}.
\newblock \DOIprefix\doi{10.1016/j.energy.2015.03.031}.
\bibitem[{Dugan and McDermott(2011)}]{dugan2011}
\bibinfo{author}{Dugan, R.C.}, \bibinfo{author}{McDermott, T.E.},
  \bibinfo{year}{2011}.
\newblock \bibinfo{title}{{An open source platform for collaborating on smart
  grid research}}, in: \bibinfo{booktitle}{2011 IEEE Power and Energy Society
  General Meeting}, pp. \bibinfo{pages}{1--7}.
\newblock \DOIprefix\doi{10.1109/PES.2011.6039829}.
\bibitem[{Duman and G{\"u}ler(2020)}]{duman2020}
\bibinfo{author}{Duman, A.C.}, \bibinfo{author}{G{\"u}ler, {\"O}.},
  \bibinfo{year}{2020}.
\newblock \bibinfo{title}{{Economic analysis of grid-connected residential
  rooftop PV systems in Turkey}}.
\newblock \bibinfo{journal}{Renewable Energy} \bibinfo{volume}{148},
  \bibinfo{pages}{697--711}.
\newblock \DOIprefix\doi{10.1016/j.renene.2019.10.157}.
\bibitem[{Ecker et~al.(2012)Ecker, Gerschler, Vogel, K{\"a}bitz, Hust, Dechent
  and Sauer}]{ecker2012}
\bibinfo{author}{Ecker, M.}, \bibinfo{author}{Gerschler, J.B.},
  \bibinfo{author}{Vogel, J.}, \bibinfo{author}{K{\"a}bitz, S.},
  \bibinfo{author}{Hust, F.}, \bibinfo{author}{Dechent, P.},
  \bibinfo{author}{Sauer, D.U.}, \bibinfo{year}{2012}.
\newblock \bibinfo{title}{{Development of a lifetime prediction model for
  lithium-ion batteries based on extended accelerated aging test data}}.
\newblock \bibinfo{journal}{Journal of Power Sources} \bibinfo{volume}{215},
  \bibinfo{pages}{248--257}.
\newblock \DOIprefix\doi{10.1016/j.jpowsour.2012.05.012}.
\bibitem[{{EPDK}(2024)}]{epdk2024}
\bibinfo{author}{{EPDK}}, \bibinfo{year}{2024}.
\newblock \bibinfo{title}{{Elektrik Piyasas{\i} 2023 Y{\i}l{\i} Piyasa
  Geli{\c{s}}im Raporu [Electricity market development report 2023]}}.
\newblock \bibinfo{type}{Technical Report}. Energy Market Regulatory Authority
  of T{\"u}rkiye (EPDK). \bibinfo{address}{Ankara}.
\newblock \URLprefix
  \url{https://www.epdk.gov.tr/Detay/DownloadDocument?id=fx+hB3E2Cns=}.
\bibitem[{{EPDK}(2026a)}]{epdk2026a}
\bibinfo{author}{{EPDK}}, \bibinfo{year}{2026}a.
\newblock \bibinfo{title}{{Elektrik Piyasas{\i}nda Lisanss{\i}z Elektrik
  {\"U}retim Y{\"o}netmeli{\u g}inde De{\u g}i{\c s}iklik Yap{\i}lmas{\i}na
  Dair Y{\"o}netmelik [Regulation amending the regulation on unlicensed
  electricity generation]}}.
\newblock \bibinfo{howpublished}{Resm{\^i} Gazete 33212, 2 April 2026}.
\newblock \URLprefix
  \url{https://www.resmigazete.gov.tr/eskiler/2026/04/20260402-4.htm}.
\bibitem[{{EPDK}(2026b)}]{epdk2026tariff}
\bibinfo{author}{{EPDK}}, \bibinfo{year}{2026}b.
\newblock \bibinfo{title}{{Enerji Piyasas{\i} D{\"u}zenleme Kurulunun
  02/04/2026 tarihli ve 14461 say{\i}l{\i} Karar{\i} [Board Decision 14461 of 2
  April 2026: electricity tariff tables in force from 4 April 2026]}}.
\newblock \bibinfo{howpublished}{Resm{\^i} Gazete 33214, 4 April 2026}.
\bibitem[{{EPDK}(2026c)}]{epdk2026b}
\bibinfo{author}{{EPDK}}, \bibinfo{year}{2026}c.
\newblock \bibinfo{title}{{Son Kaynak Tedarik Tarifesi (SKTT): s{\i}k{\c c}a
  sorulan sorular [Last-resort supply tariff: frequently asked questions]}}.
\newblock \URLprefix
  \url{https://www.epdk.gov.tr/Detay/Icerik/16-38/son-kaynak-tedarik-tarifesi-sktt-sikca-sorulan-}.
  \bibinfo{note}{accessed 15 November 2026}.
\bibitem[{{EP{\.I}A{\c S}}(2026a)}]{epias2026a}
\bibinfo{author}{{EP{\.I}A{\c S}}}, \bibinfo{year}{2026}a.
\newblock \bibinfo{title}{{{\c S}effafl{\i}k Platformu: Ger{\c c}ek Zamanl{\i}
  T{\"u}ketim [Transparency platform: real-time consumption], hourly, 2023}}.
\newblock \URLprefix \url{https://seffaflik.epias.com.tr}.
\bibitem[{{EP{\.I}A{\c S}}(2026b)}]{epias2026b}
\bibinfo{author}{{EP{\.I}A{\c S}}}, \bibinfo{year}{2026}b.
\newblock \bibinfo{title}{{{\c S}effafl{\i}k Platformu: YEKDEM birim maliyeti
  ve Piyasa Takas Fiyat{\i} [Transparency platform: YEKDEM unit cost and market
  clearing price], monthly, October 2025 -- September 2026}}.
\newblock \URLprefix \url{https://seffaflik.epias.com.tr}.
\bibitem[{Freeman et~al.(2014)Freeman, Whitmore, Blair and Dobos}]{freeman2014}
\bibinfo{author}{Freeman, J.}, \bibinfo{author}{Whitmore, J.},
  \bibinfo{author}{Blair, N.}, \bibinfo{author}{Dobos, A.P.},
  \bibinfo{year}{2014}.
\newblock \bibinfo{title}{{Validation of Multiple Tools for Flat Plate
  Photovoltaic Modeling Against Measured Data}}.
\newblock \bibinfo{type}{Technical Report}
  \bibinfo{number}{NREL/TP-6A20-61497}. National Renewable Energy Laboratory.
  \bibinfo{address}{Golden, CO}.
\newblock \DOIprefix\doi{10.2172/1150179}.
\bibitem[{F{\"u}rst et~al.(2020)F{\"u}rst, Chen, Gu and Tong}]{furst2020}
\bibinfo{author}{F{\"u}rst, K.}, \bibinfo{author}{Chen, P.},
  \bibinfo{author}{Gu, I.Y.H.}, \bibinfo{author}{Tong, L.},
  \bibinfo{year}{2020}.
\newblock \bibinfo{title}{{Improved peak load estimation from single and
  multiple consumer categories}}.
\newblock \bibinfo{journal}{CIRED - Open Access Proceedings Journal}
  \bibinfo{volume}{2020}, \bibinfo{pages}{178--181}.
\newblock \DOIprefix\doi{10.1049/oap-cired.2021.0300}.
\bibitem[{Guillaume et~al.(2016)Guillaume, Arshad, Jakeman, Jalava and
  Kummu}]{guillaume2016}
\bibinfo{author}{Guillaume, J.H.}, \bibinfo{author}{Arshad, M.},
  \bibinfo{author}{Jakeman, A.J.}, \bibinfo{author}{Jalava, M.},
  \bibinfo{author}{Kummu, M.}, \bibinfo{year}{2016}.
\newblock \bibinfo{title}{{Robust discrimination between uncertain management
  alternatives by iterative reflection on crossover point scenarios:
  Principles, design and implementations}}.
\newblock \bibinfo{journal}{Environmental Modelling \& Software}
  \bibinfo{volume}{83}, \bibinfo{pages}{326--343}.
\newblock \DOIprefix\doi{10.1016/j.envsoft.2016.04.005}.
\bibitem[{Hay and Davies(1980)}]{hay1980}
\bibinfo{author}{Hay, J.E.}, \bibinfo{author}{Davies, J.A.},
  \bibinfo{year}{1980}.
\newblock \bibinfo{title}{{Calculations of the solar radiation incident on an
  inclined surface}}, in: \bibinfo{editor}{Hay, J.E.}, \bibinfo{editor}{Won,
  T.K.} (Eds.), \bibinfo{booktitle}{Proceedings of the First Canadian Solar
  Radiation Data Workshop, Toronto, 17--19 April 1978},
  \bibinfo{publisher}{Supply and Services Canada}, \bibinfo{address}{Ottawa}.
  pp. \bibinfo{pages}{59--72}.
\bibitem[{Herman and Usher(2017)}]{herman2017}
\bibinfo{author}{Herman, J.}, \bibinfo{author}{Usher, W.},
  \bibinfo{year}{2017}.
\newblock \bibinfo{title}{{SALib: An open-source Python library for Sensitivity
  Analysis}}.
\newblock \bibinfo{journal}{The Journal of Open Source Software}
  \bibinfo{volume}{2}, \bibinfo{pages}{97}.
\newblock \DOIprefix\doi{10.21105/joss.00097}.
\bibitem[{Higgins et~al.(2014)Higgins, McNamara and Foliente}]{higgins2014}
\bibinfo{author}{Higgins, A.}, \bibinfo{author}{McNamara, C.},
  \bibinfo{author}{Foliente, G.}, \bibinfo{year}{2014}.
\newblock \bibinfo{title}{{Modelling future uptake of solar photo-voltaics and
  water heaters under different government incentives}}.
\newblock \bibinfo{journal}{Technological Forecasting and Social Change}
  \bibinfo{volume}{83}, \bibinfo{pages}{142--155}.
\newblock \DOIprefix\doi{10.1016/j.techfore.2013.07.006}.
\bibitem[{{HOMER Energy}(2026)}]{homer2026}
\bibinfo{author}{{HOMER Energy}}, \bibinfo{year}{2026}.
\newblock \bibinfo{title}{{HOMER Pro and HOMER Grid user manuals [online
  help]}}.
\newblock \URLprefix \url{https://homerenergy.com/products/pro/docs/latest/}.
  \bibinfo{note}{accessed 15 November 2026}.
\bibitem[{Huang and Ma(2024)}]{huang2024}
\bibinfo{author}{Huang, P.}, \bibinfo{author}{Ma, Z.}, \bibinfo{year}{2024}.
\newblock \bibinfo{title}{{Unveiling electric vehicle (EV) charging patterns
  and their transformative role in electricity balancing and delivery: Insights
  from real-world data in Sweden}}.
\newblock \bibinfo{journal}{Renewable Energy} \bibinfo{volume}{236},
  \bibinfo{pages}{121511}.
\newblock \DOIprefix\doi{10.1016/j.renene.2024.121511}.
\bibitem[{Huld et~al.(2012)Huld, M{\"u}ller and Gambardella}]{huld2012}
\bibinfo{author}{Huld, T.}, \bibinfo{author}{M{\"u}ller, R.},
  \bibinfo{author}{Gambardella, A.}, \bibinfo{year}{2012}.
\newblock \bibinfo{title}{{A new solar radiation database for estimating PV
  performance in Europe and Africa}}.
\newblock \bibinfo{journal}{Solar Energy} \bibinfo{volume}{86},
  \bibinfo{pages}{1803--1815}.
\newblock \DOIprefix\doi{10.1016/j.solener.2012.03.006}.
\bibitem[{Huppmann et~al.(2019)Huppmann, Gidden, Fricko, Kolp, Orthofer,
  Pimmer, Kushin, Vinca, Mastrucci, Riahi and Krey}]{huppmann2019}
\bibinfo{author}{Huppmann, D.}, \bibinfo{author}{Gidden, M.},
  \bibinfo{author}{Fricko, O.}, \bibinfo{author}{Kolp, P.},
  \bibinfo{author}{Orthofer, C.}, \bibinfo{author}{Pimmer, M.},
  \bibinfo{author}{Kushin, N.}, \bibinfo{author}{Vinca, A.},
  \bibinfo{author}{Mastrucci, A.}, \bibinfo{author}{Riahi, K.},
  \bibinfo{author}{Krey, V.}, \bibinfo{year}{2019}.
\newblock \bibinfo{title}{{The MESSAGE Integrated Assessment Model and the ix
  modeling platform (ixmp): An open framework for integrated and cross-cutting
  analysis of energy, climate, the environment, and sustainable development}}.
\newblock \bibinfo{journal}{Environmental Modelling \& Software}
  \bibinfo{volume}{112}, \bibinfo{pages}{143--156}.
\newblock \DOIprefix\doi{10.1016/j.envsoft.2018.11.012}.
\bibitem[{Kaynak(2026)}]{kaynak2026}
\bibinfo{author}{Kaynak, B.}, \bibinfo{year}{2026}.
\newblock \bibinfo{title}{{Right Number, Wrong Scale: Measuring Grounding
  Fidelity in {LLM}-Orchestrated Energy-Investment Decision Support from
  Buildings to Feeders}}.
\newblock \bibinfo{note}{Manuscript in preparation}.
\bibitem[{King et~al.(2004)King, Boyson and Kratochvil}]{king2004}
\bibinfo{author}{King, D.L.}, \bibinfo{author}{Boyson, W.E.},
  \bibinfo{author}{Kratochvil, J.A.}, \bibinfo{year}{2004}.
\newblock \bibinfo{title}{{Photovoltaic Array Performance Model}}.
\newblock \bibinfo{type}{Technical Report} \bibinfo{number}{SAND2004-3535}.
  Sandia National Laboratories. \bibinfo{address}{Albuquerque, NM}.
\newblock \DOIprefix\doi{10.2172/919131}.
\bibitem[{Krishnan et~al.(2020)Krishnan, Bugbee, Elgindy, Mateo, Duenas,
  Postigo, Lacroix, G{\'o}mez San~Rom{\'a}n and Palmintier}]{krishnan2020}
\bibinfo{author}{Krishnan, V.}, \bibinfo{author}{Bugbee, B.},
  \bibinfo{author}{Elgindy, T.}, \bibinfo{author}{Mateo, C.},
  \bibinfo{author}{Duenas, P.}, \bibinfo{author}{Postigo, F.},
  \bibinfo{author}{Lacroix, J.S.}, \bibinfo{author}{G{\'o}mez San~Rom{\'a}n,
  T.}, \bibinfo{author}{Palmintier, B.}, \bibinfo{year}{2020}.
\newblock \bibinfo{title}{{Validation of Synthetic U.S. Electric Power
  Distribution System Data Sets}}.
\newblock \bibinfo{journal}{IEEE Transactions on Smart Grid}
  \bibinfo{volume}{11}, \bibinfo{pages}{4477--4489}.
\newblock \DOIprefix\doi{10.1109/TSG.2020.2981077}.
\bibitem[{Li et~al.(2025)Li, Xu and Hong}]{li2025}
\bibinfo{author}{Li, H.}, \bibinfo{author}{Xu, Y.}, \bibinfo{author}{Hong, T.},
  \bibinfo{year}{2025}.
\newblock \bibinfo{title}{{EnergyPlus-MCP: A model-context-protocol server for
  AI-driven building energy modeling}}.
\newblock \bibinfo{journal}{SoftwareX} \bibinfo{volume}{32},
  \bibinfo{pages}{102367}.
\newblock \DOIprefix\doi{10.1016/j.softx.2025.102367}.
\bibitem[{Liu et~al.(2026)Liu, Dong and Lian}]{liu2026}
\bibinfo{author}{Liu, B.}, \bibinfo{author}{Dong, J.}, \bibinfo{author}{Lian,
  J.}, \bibinfo{year}{2026}.
\newblock \bibinfo{title}{{Grid-Orch: An LLM-Powered Orchestrator for
  Distribution Grid Simulation and Analytics}}.
\newblock \bibinfo{journal}{IEEE Open Access Journal of Power and Energy}
  \bibinfo{volume}{13}, \bibinfo{pages}{727--738}.
\newblock \DOIprefix\doi{10.1109/OAJPE.2026.3721083}.
\bibitem[{Londo et~al.(2020)Londo, Matton, Usmani, van Klaveren, Tigchelaar and
  Brunsting}]{londo2020}
\bibinfo{author}{Londo, M.}, \bibinfo{author}{Matton, R.},
  \bibinfo{author}{Usmani, O.}, \bibinfo{author}{van Klaveren, M.},
  \bibinfo{author}{Tigchelaar, C.}, \bibinfo{author}{Brunsting, S.},
  \bibinfo{year}{2020}.
\newblock \bibinfo{title}{{Alternatives for current net metering policy for
  solar PV in the Netherlands: A comparison of impacts on business case and
  purchasing behaviour of private homeowners, and on governmental costs}}.
\newblock \bibinfo{journal}{Renewable Energy} \bibinfo{volume}{147},
  \bibinfo{pages}{903--915}.
\newblock \DOIprefix\doi{10.1016/j.renene.2019.09.062}.
\bibitem[{Lyu and Rozhkov(2026)}]{lyu2026}
\bibinfo{author}{Lyu, S.}, \bibinfo{author}{Rozhkov, A.}, \bibinfo{year}{2026}.
\newblock \bibinfo{title}{{Where incentives hit the roof: How policy design
  shapes rooftop solar returns}}.
\newblock \bibinfo{journal}{Renewable Energy} \bibinfo{volume}{273},
  \bibinfo{pages}{126123}.
\newblock \DOIprefix\doi{10.1016/j.renene.2026.126123}.
\bibitem[{Maier et~al.(2016)Maier, Guillaume, van Delden, Riddell, Haasnoot and
  Kwakkel}]{maier2016}
\bibinfo{author}{Maier, H.}, \bibinfo{author}{Guillaume, J.},
  \bibinfo{author}{van Delden, H.}, \bibinfo{author}{Riddell, G.},
  \bibinfo{author}{Haasnoot, M.}, \bibinfo{author}{Kwakkel, J.},
  \bibinfo{year}{2016}.
\newblock \bibinfo{title}{{An uncertain future, deep uncertainty, scenarios,
  robustness and adaptation: How do they fit together?}}
\newblock \bibinfo{journal}{Environmental Modelling \& Software}
  \bibinfo{volume}{81}, \bibinfo{pages}{154--164}.
\newblock \DOIprefix\doi{10.1016/j.envsoft.2016.03.014}.
\bibitem[{Meier et~al.(1999)Meier, F{\"u}nfgeld, Adam and
  Schieferdecker}]{meier1999}
\bibinfo{author}{Meier, H.}, \bibinfo{author}{F{\"u}nfgeld, C.},
  \bibinfo{author}{Adam, T.}, \bibinfo{author}{Schieferdecker, B.},
  \bibinfo{year}{1999}.
\newblock \bibinfo{title}{{Repr{\"a}sentative VDEW-Lastprofile}}.
\newblock \bibinfo{type}{Technical Report} \bibinfo{number}{VDEW-Materialien
  M-32/99}. VDEW. \bibinfo{address}{Frankfurt am Main}.
\newblock \URLprefix
  \url{https://www.bdew.de/media/documents/1999_Repraesentative-VDEW-Lastprofile.pdf}.
\bibitem[{Meinecke et~al.(2020)Meinecke, Sarajli{\'c}, Drauz, Klettke, Lauven,
  Rehtanz, Moser and Braun}]{meinecke2020}
\bibinfo{author}{Meinecke, S.}, \bibinfo{author}{Sarajli{\'c}, D.},
  \bibinfo{author}{Drauz, S.R.}, \bibinfo{author}{Klettke, A.},
  \bibinfo{author}{Lauven, L.P.}, \bibinfo{author}{Rehtanz, C.},
  \bibinfo{author}{Moser, A.}, \bibinfo{author}{Braun, M.},
  \bibinfo{year}{2020}.
\newblock \bibinfo{title}{{SimBench---A Benchmark Dataset of Electric Power
  Systems to Compare Innovative Solutions Based on Power Flow Analysis}}.
\newblock \bibinfo{journal}{Energies} \bibinfo{volume}{13},
  \bibinfo{pages}{3290}.
\newblock \DOIprefix\doi{10.3390/en13123290}.
\bibitem[{Mishra et~al.(2022)Mishra, Pohl, Laws, Cutler, Kwasnik, Becker,
  Zolan, Anderson, Olis and Elgqvist}]{mishra2022}
\bibinfo{author}{Mishra, S.}, \bibinfo{author}{Pohl, J.},
  \bibinfo{author}{Laws, N.}, \bibinfo{author}{Cutler, D.},
  \bibinfo{author}{Kwasnik, T.}, \bibinfo{author}{Becker, W.},
  \bibinfo{author}{Zolan, A.}, \bibinfo{author}{Anderson, K.},
  \bibinfo{author}{Olis, D.}, \bibinfo{author}{Elgqvist, E.},
  \bibinfo{year}{2022}.
\newblock \bibinfo{title}{{Computational framework for behind-the-meter DER
  techno-economic modeling and optimization: REopt Lite}}.
\newblock \bibinfo{journal}{Energy Systems} \bibinfo{volume}{13},
  \bibinfo{pages}{509--537}.
\newblock \DOIprefix\doi{10.1007/s12667-021-00446-8}.
\bibitem[{Morris(1991)}]{morris1991}
\bibinfo{author}{Morris, M.D.}, \bibinfo{year}{1991}.
\newblock \bibinfo{title}{{Factorial Sampling Plans for Preliminary
  Computational Experiments}}.
\newblock \bibinfo{journal}{Technometrics} \bibinfo{volume}{33},
  \bibinfo{pages}{161--174}.
\newblock \DOIprefix\doi{10.1080/00401706.1991.10484804}.
\bibitem[{Mylonas et~al.(2026)Mylonas, Georgoulakis and Foti}]{mylonas2026}
\bibinfo{author}{Mylonas, C.}, \bibinfo{author}{Georgoulakis, T.},
  \bibinfo{author}{Foti, M.}, \bibinfo{year}{2026}.
\newblock \bibinfo{title}{{A Conversational Agentic Interface to Physics-Based
  Household Digital Twins for Residential Energy Decision Support}}.
\newblock \bibinfo{howpublished}{arXiv preprint arXiv:2606.31744}.
\newblock \DOIprefix\doi{10.48550/arXiv.2606.31744}.
\bibitem[{Navarro-Espinosa and Ochoa(2016)}]{navarro2016}
\bibinfo{author}{Navarro-Espinosa, A.}, \bibinfo{author}{Ochoa, L.F.},
  \bibinfo{year}{2016}.
\newblock \bibinfo{title}{{Probabilistic Impact Assessment of Low Carbon
  Technologies in LV Distribution Systems}}.
\newblock \bibinfo{journal}{IEEE Transactions on Power Systems}
  \bibinfo{volume}{31}, \bibinfo{pages}{2192--2203}.
\newblock \DOIprefix\doi{10.1109/TPWRS.2015.2448663}.
\bibitem[{{NREL}(2026)}]{urbanopt2026}
\bibinfo{author}{{NREL}}, \bibinfo{year}{2026}.
\newblock \bibinfo{title}{{{URBANopt} documentation: {OpenDSS} and {RNM-US}
  workflows}}.
\newblock \URLprefix \url{https://docs.urbanopt.net/}. \bibinfo{note}{accessed
  18 November 2026}.
\bibitem[{{oemof developer group}(2023)}]{oemof2023}
\bibinfo{author}{{oemof developer group}}, \bibinfo{year}{2023}.
\newblock \bibinfo{title}{{demandlib: Demandlib of the open energy modelling
  framework, Version 0.1.9}}.
\newblock \bibinfo{howpublished}{Software}.
\newblock \URLprefix \url{https://pypi.org/project/demandlib/0.1.9/}.
  \bibinfo{note}{{DOI} of all versions: 10.5281/zenodo.2553504}.
\bibitem[{{OpenStreetMap contributors}(2026)}]{osm2026}
\bibinfo{author}{{OpenStreetMap contributors}}, \bibinfo{year}{2026}.
\newblock \bibinfo{title}{{OpenStreetMap}}.
\newblock \URLprefix \url{https://www.openstreetmap.org}.
\bibitem[{Pianosi et~al.(2016)Pianosi, Beven, Freer, Hall, Rougier, Stephenson
  and Wagener}]{pianosi2016}
\bibinfo{author}{Pianosi, F.}, \bibinfo{author}{Beven, K.},
  \bibinfo{author}{Freer, J.}, \bibinfo{author}{Hall, J.W.},
  \bibinfo{author}{Rougier, J.}, \bibinfo{author}{Stephenson, D.B.},
  \bibinfo{author}{Wagener, T.}, \bibinfo{year}{2016}.
\newblock \bibinfo{title}{{Sensitivity analysis of environmental models: A
  systematic review with practical workflow}}.
\newblock \bibinfo{journal}{Environmental Modelling \& Software}
  \bibinfo{volume}{79}, \bibinfo{pages}{214--232}.
\newblock \DOIprefix\doi{10.1016/j.envsoft.2016.02.008}.
\bibitem[{{Presidency of T{\"u}rkiye}(2026)}]{presidency2026}
\bibinfo{author}{{Presidency of T{\"u}rkiye}}, \bibinfo{year}{2026}.
\newblock \bibinfo{title}{{Cumhurba{\c s}kan{\i} Karar{\i} 11415 [Presidential
  Decision 11415 on the price of excess energy after the ten-year support
  period]}}.
\newblock \bibinfo{howpublished}{Resm{\^i} Gazete 33279, 13 June 2026}.
\newblock \URLprefix
  \url{https://www.resmigazete.gov.tr/eskiler/2026/06/20260613-3.pdf}.
\bibitem[{Ravalico et~al.(2010)Ravalico, Dandy and Maier}]{ravalico2010}
\bibinfo{author}{Ravalico, J.}, \bibinfo{author}{Dandy, G.},
  \bibinfo{author}{Maier, H.}, \bibinfo{year}{2010}.
\newblock \bibinfo{title}{{Management Option Rank Equivalence (MORE) -- A new
  method of sensitivity analysis for decision-making}}.
\newblock \bibinfo{journal}{Environmental Modelling \& Software}
  \bibinfo{volume}{25}, \bibinfo{pages}{171--181}.
\newblock \DOIprefix\doi{10.1016/j.envsoft.2009.06.012}.
\bibitem[{Razavi et~al.(2021)Razavi, Jakeman, Saltelli, Prieur, Iooss,
  Borgonovo, Plischke, Lo~Piano, Iwanaga, Becker, Tarantola, Guillaume,
  Jakeman, Gupta, Melillo, Rabitti, Chabridon, Duan, Sun, Smith,
  Sheikholeslami, Hosseini, Asadzadeh, Puy, Kucherenko and Maier}]{razavi2021}
\bibinfo{author}{Razavi, S.}, \bibinfo{author}{Jakeman, A.},
  \bibinfo{author}{Saltelli, A.}, \bibinfo{author}{Prieur, C.},
  \bibinfo{author}{Iooss, B.}, \bibinfo{author}{Borgonovo, E.},
  \bibinfo{author}{Plischke, E.}, \bibinfo{author}{Lo~Piano, S.},
  \bibinfo{author}{Iwanaga, T.}, \bibinfo{author}{Becker, W.},
  \bibinfo{author}{Tarantola, S.}, \bibinfo{author}{Guillaume, J.H.},
  \bibinfo{author}{Jakeman, J.}, \bibinfo{author}{Gupta, H.},
  \bibinfo{author}{Melillo, N.}, \bibinfo{author}{Rabitti, G.},
  \bibinfo{author}{Chabridon, V.}, \bibinfo{author}{Duan, Q.},
  \bibinfo{author}{Sun, X.}, \bibinfo{author}{Smith, S.},
  \bibinfo{author}{Sheikholeslami, R.}, \bibinfo{author}{Hosseini, N.},
  \bibinfo{author}{Asadzadeh, M.}, \bibinfo{author}{Puy, A.},
  \bibinfo{author}{Kucherenko, S.}, \bibinfo{author}{Maier, H.R.},
  \bibinfo{year}{2021}.
\newblock \bibinfo{title}{{The Future of Sensitivity Analysis: An essential
  discipline for systems modeling and policy support}}.
\newblock \bibinfo{journal}{Environmental Modelling \& Software}
  \bibinfo{volume}{137}, \bibinfo{pages}{104954}.
\newblock \DOIprefix\doi{10.1016/j.envsoft.2020.104954}.
\bibitem[{Reveron~Baecker et~al.(2025)Reveron~Baecker, Candas, Tepe and
  Mohapatra}]{reveron2025}
\bibinfo{author}{Reveron~Baecker, B.}, \bibinfo{author}{Candas, S.},
  \bibinfo{author}{Tepe, D.}, \bibinfo{author}{Mohapatra, A.},
  \bibinfo{year}{2025}.
\newblock \bibinfo{title}{{Generation of low-voltage synthetic grid data for
  energy system modeling with the pylovo tool}}.
\newblock \bibinfo{journal}{Sustainable Energy, Grids and Networks}
  \bibinfo{volume}{41}, \bibinfo{pages}{101617}.
\newblock \DOIprefix\doi{10.1016/j.segan.2024.101617}.
\bibitem[{Saltelli et~al.(2019)Saltelli, Aleksankina, Becker, Fennell,
  Ferretti, Holst, Li and Wu}]{saltelli2019}
\bibinfo{author}{Saltelli, A.}, \bibinfo{author}{Aleksankina, K.},
  \bibinfo{author}{Becker, W.}, \bibinfo{author}{Fennell, P.},
  \bibinfo{author}{Ferretti, F.}, \bibinfo{author}{Holst, N.},
  \bibinfo{author}{Li, S.}, \bibinfo{author}{Wu, Q.}, \bibinfo{year}{2019}.
\newblock \bibinfo{title}{{Why so many published sensitivity analyses are
  false: A systematic review of sensitivity analysis practices}}.
\newblock \bibinfo{journal}{Environmental Modelling \& Software}
  \bibinfo{volume}{114}, \bibinfo{pages}{29--39}.
\newblock \DOIprefix\doi{10.1016/j.envsoft.2019.01.012}.
\bibitem[{Saltelli et~al.(2010)Saltelli, Annoni, Azzini, Campolongo, Ratto and
  Tarantola}]{saltelli2010}
\bibinfo{author}{Saltelli, A.}, \bibinfo{author}{Annoni, P.},
  \bibinfo{author}{Azzini, I.}, \bibinfo{author}{Campolongo, F.},
  \bibinfo{author}{Ratto, M.}, \bibinfo{author}{Tarantola, S.},
  \bibinfo{year}{2010}.
\newblock \bibinfo{title}{{Variance based sensitivity analysis of model output.
  Design and estimator for the total sensitivity index}}.
\newblock \bibinfo{journal}{Computer Physics Communications}
  \bibinfo{volume}{181}, \bibinfo{pages}{259--270}.
\newblock \DOIprefix\doi{10.1016/j.cpc.2009.09.018}.
\bibitem[{Schmalstieg et~al.(2014)Schmalstieg, K{\"a}bitz, Ecker and
  Sauer}]{schmalstieg2014}
\bibinfo{author}{Schmalstieg, J.}, \bibinfo{author}{K{\"a}bitz, S.},
  \bibinfo{author}{Ecker, M.}, \bibinfo{author}{Sauer, D.U.},
  \bibinfo{year}{2014}.
\newblock \bibinfo{title}{{A holistic aging model for Li(NiMnCo)O2 based 18650
  lithium-ion batteries}}.
\newblock \bibinfo{journal}{Journal of Power Sources} \bibinfo{volume}{257},
  \bibinfo{pages}{325--334}.
\newblock \DOIprefix\doi{10.1016/j.jpowsour.2014.02.012}.
\bibitem[{Sigrin et~al.(2016)Sigrin, Gleason, Preus, Baring-Gould and
  Margolis}]{sigrin2016}
\bibinfo{author}{Sigrin, B.}, \bibinfo{author}{Gleason, M.},
  \bibinfo{author}{Preus, R.}, \bibinfo{author}{Baring-Gould, I.},
  \bibinfo{author}{Margolis, R.}, \bibinfo{year}{2016}.
\newblock \bibinfo{title}{{The Distributed Generation Market Demand Model
  (dGen): Documentation}}.
\newblock \bibinfo{type}{Technical Report}
  \bibinfo{number}{NREL/TP-6A20-65231}. National Renewable Energy Laboratory.
  \bibinfo{address}{Golden, CO}.
\newblock \DOIprefix\doi{10.2172/1239054}.
\bibitem[{Sirko et~al.(2021)Sirko, Kashubin, Ritter, Annkah, Bouchareb,
  Dauphin, Keysers, Neumann, Cisse and Quinn}]{sirko2021}
\bibinfo{author}{Sirko, W.}, \bibinfo{author}{Kashubin, S.},
  \bibinfo{author}{Ritter, M.}, \bibinfo{author}{Annkah, A.},
  \bibinfo{author}{Bouchareb, Y.S.E.}, \bibinfo{author}{Dauphin, Y.},
  \bibinfo{author}{Keysers, D.}, \bibinfo{author}{Neumann, M.},
  \bibinfo{author}{Cisse, M.}, \bibinfo{author}{Quinn, J.},
  \bibinfo{year}{2021}.
\newblock \bibinfo{title}{{Continental-Scale Building Detection from High
  Resolution Satellite Imagery}}.
\newblock \bibinfo{howpublished}{arXiv preprint arXiv:2107.12283}.
\newblock \DOIprefix\doi{10.48550/arXiv.2107.12283}.
\bibitem[{Spear and Hornberger(1980)}]{spear1980}
\bibinfo{author}{Spear, R.}, \bibinfo{author}{Hornberger, G.},
  \bibinfo{year}{1980}.
\newblock \bibinfo{title}{{Eutrophication in peel inlet---II. Identification of
  critical uncertainties via generalized sensitivity analysis}}.
\newblock \bibinfo{journal}{Water Research} \bibinfo{volume}{14},
  \bibinfo{pages}{43--49}.
\newblock \DOIprefix\doi{10.1016/0043-1354(80)90040-8}.
\bibitem[{Su et~al.(2020)Su, Kern, Denaro, Hill, Reed, Sun, Cohen and
  Characklis}]{su2020}
\bibinfo{author}{Su, Y.}, \bibinfo{author}{Kern, J.D.},
  \bibinfo{author}{Denaro, S.}, \bibinfo{author}{Hill, J.},
  \bibinfo{author}{Reed, P.}, \bibinfo{author}{Sun, Y.},
  \bibinfo{author}{Cohen, J.}, \bibinfo{author}{Characklis, G.W.},
  \bibinfo{year}{2020}.
\newblock \bibinfo{title}{{An open source model for quantifying risks in bulk
  electric power systems from spatially and temporally correlated
  hydrometeorological processes}}.
\newblock \bibinfo{journal}{Environmental Modelling \& Software}
  \bibinfo{volume}{126}, \bibinfo{pages}{104667}.
\newblock \DOIprefix\doi{10.1016/j.envsoft.2020.104667}.
\bibitem[{Tantisattayakul and Kanchanapiya(2026)}]{tantisattayakul2026}
\bibinfo{author}{Tantisattayakul, T.}, \bibinfo{author}{Kanchanapiya, P.},
  \bibinfo{year}{2026}.
\newblock \bibinfo{title}{{Adoption, fiscal cost, and policy compromise:
  Evaluating residential rooftop solar policies in Thailand for emerging
  economies}}.
\newblock \bibinfo{journal}{Energy for Sustainable Development}
  \bibinfo{volume}{93}, \bibinfo{pages}{101986}.
\newblock \DOIprefix\doi{10.1016/j.esd.2026.101986}.
\bibitem[{Thurner et~al.(2018)Thurner, Scheidler, Sch{\"a}fer, Menke,
  Dollichon, Meier, Meinecke and Braun}]{thurner2018}
\bibinfo{author}{Thurner, L.}, \bibinfo{author}{Scheidler, A.},
  \bibinfo{author}{Sch{\"a}fer, F.}, \bibinfo{author}{Menke, J.H.},
  \bibinfo{author}{Dollichon, J.}, \bibinfo{author}{Meier, F.},
  \bibinfo{author}{Meinecke, S.}, \bibinfo{author}{Braun, M.},
  \bibinfo{year}{2018}.
\newblock \bibinfo{title}{{Pandapower---An Open-Source Python Tool for
  Convenient Modeling, Analysis, and Optimization of Electric Power Systems}}.
\newblock \bibinfo{journal}{IEEE Transactions on Power Systems}
  \bibinfo{volume}{33}, \bibinfo{pages}{6510--6521}.
\newblock \DOIprefix\doi{10.1109/TPWRS.2018.2829021}.
\bibitem[{Vaishnav et~al.(2017)Vaishnav, Horner and Azevedo}]{vaishnav2017}
\bibinfo{author}{Vaishnav, P.}, \bibinfo{author}{Horner, N.},
  \bibinfo{author}{Azevedo, I.L.}, \bibinfo{year}{2017}.
\newblock \bibinfo{title}{{Was it worthwhile? Where have the benefits of
  rooftop solar photovoltaic generation exceeded the cost?}}
\newblock \bibinfo{journal}{Environmental Research Letters}
  \bibinfo{volume}{12}, \bibinfo{pages}{094015}.
\newblock \DOIprefix\doi{10.1088/1748-9326/aa815e}.
\bibitem[{Vimpari and Junnila(2019)}]{vimpari2019}
\bibinfo{author}{Vimpari, J.}, \bibinfo{author}{Junnila, S.},
  \bibinfo{year}{2019}.
\newblock \bibinfo{title}{{Estimating the diffusion of rooftop PVs: A real
  estate economics perspective}}.
\newblock \bibinfo{journal}{Energy} \bibinfo{volume}{172},
  \bibinfo{pages}{1087--1097}.
\newblock \DOIprefix\doi{10.1016/j.energy.2019.02.049}.
\bibitem[{Voronin et~al.(2025)Voronin, Nepsha and Ilyushin}]{voronin2025}
\bibinfo{author}{Voronin, V.}, \bibinfo{author}{Nepsha, F.},
  \bibinfo{author}{Ilyushin, P.}, \bibinfo{year}{2025}.
\newblock \bibinfo{title}{{A Demand Factor Analysis for Electric Vehicle
  Charging Infrastructure}}.
\newblock \bibinfo{journal}{World Electric Vehicle Journal}
  \bibinfo{volume}{16}, \bibinfo{pages}{537}.
\newblock \DOIprefix\doi{10.3390/wevj16090537}.
\bibitem[{Weber et~al.(2024)Weber, Janecke, {\c C}akmak and
  Hagenmeyer}]{weber2024}
\bibinfo{author}{Weber, M.}, \bibinfo{author}{Janecke, L.},
  \bibinfo{author}{{\c C}akmak, H.K.}, \bibinfo{author}{Hagenmeyer, V.},
  \bibinfo{year}{2024}.
\newblock \bibinfo{title}{{Open Data-Driven Automation of Residential
  Distribution Grid Modeling With Minimal Data Requirements}}.
\newblock \bibinfo{journal}{IEEE Transactions on Smart Grid}
  \bibinfo{volume}{15}, \bibinfo{pages}{5721--5732}.
\newblock \DOIprefix\doi{10.1109/TSG.2024.3406765}.
\bibitem[{Wilson(1927)}]{wilson1927}
\bibinfo{author}{Wilson, E.B.}, \bibinfo{year}{1927}.
\newblock \bibinfo{title}{{Probable Inference, the Law of Succession, and
  Statistical Inference}}.
\newblock \bibinfo{journal}{Journal of the American Statistical Association}
  \bibinfo{volume}{22}, \bibinfo{pages}{209--212}.
\newblock \DOIprefix\doi{10.1080/01621459.1927.10502953}.
\bibitem[{Xexakis and Trutnevyte(2019)}]{xexakis2019}
\bibinfo{author}{Xexakis, G.}, \bibinfo{author}{Trutnevyte, E.},
  \bibinfo{year}{2019}.
\newblock \bibinfo{title}{{Are interactive web-tools for environmental scenario
  visualization worth the effort? An experimental study on the Swiss
  electricity supply scenarios 2035}}.
\newblock \bibinfo{journal}{Environmental Modelling \& Software}
  \bibinfo{volume}{119}, \bibinfo{pages}{124--134}.
\newblock \DOIprefix\doi{10.1016/j.envsoft.2019.05.014}.
\bibitem[{Zakeri et~al.(2021)Zakeri, Cross, Dodds and Gissey}]{zakeri2021}
\bibinfo{author}{Zakeri, B.}, \bibinfo{author}{Cross, S.},
  \bibinfo{author}{Dodds, P.}, \bibinfo{author}{Gissey, G.C.},
  \bibinfo{year}{2021}.
\newblock \bibinfo{title}{{Policy options for enhancing economic profitability
  of residential solar photovoltaic with battery energy storage}}.
\newblock \bibinfo{journal}{Applied Energy} \bibinfo{volume}{290},
  \bibinfo{pages}{116697}.
\newblock \DOIprefix\doi{10.1016/j.apenergy.2021.116697}.
\bibitem[{Zhang and Xie(2025)}]{zhang2025}
\bibinfo{author}{Zhang, Q.}, \bibinfo{author}{Xie, L.}, \bibinfo{year}{2025}.
\newblock \bibinfo{title}{{PowerAgent: A Road Map Toward Agentic Intelligence
  in Power Systems: Foundation Model, Model Context Protocol, and Workflow}}.
\newblock \bibinfo{journal}{IEEE Power and Energy Magazine}
  \bibinfo{volume}{23}, \bibinfo{pages}{93--101}.
\newblock \DOIprefix\doi{10.1109/MPE.2025.3579718}.
\bibitem[{Zhao et~al.(2026)Zhao, Hou, Zhan, Wang, Xie, Guo, Chen, Li and
  Xue}]{zhao2026}
\bibinfo{author}{Zhao, S.}, \bibinfo{author}{Hou, Q.}, \bibinfo{author}{Zhan,
  Z.}, \bibinfo{author}{Wang, Y.}, \bibinfo{author}{Xie, Y.},
  \bibinfo{author}{Guo, Y.}, \bibinfo{author}{Chen, L.}, \bibinfo{author}{Li,
  S.}, \bibinfo{author}{Xue, Z.}, \bibinfo{year}{2026}.
\newblock \bibinfo{title}{{Parasites in the Toolchain: A Large-Scale Analysis
  of Attacks on the MCP Ecosystem}}, in: \bibinfo{booktitle}{2026 IEEE
  Symposium on Security and Privacy (SP)}, pp. \bibinfo{pages}{138--155}.
\newblock \DOIprefix\doi{10.1109/SP63933.2026.00154}.

\end{thebibliography}
\end{document}